\documentclass{aastex63}
\usepackage{xcolor}
\usepackage{longtable}
\usepackage{amsmath,amssymb}
\usepackage{rotating}
\usepackage{amssymb}
\usepackage{url}

\makeatletter

\newcommand{\Rmnum}[1]{\expandafter\@slowromancap\romannumeral #1@}
\makeatother
\usepackage{color}
\submitjournal{ApJ}

\shorttitle{Phase resolved spectrum of 1A~0535+262}
\shortauthors{Kong et al.}

\begin{document}

\title{Phase dependent evolution within large luminosity range of 1A~0535+262 observed by Insight-HXMT during 2020 giant outburst}

\author[0000-0003-3188-9079]{Ling-Da Kong\textsuperscript{*}}
\email{kongld@ihep.ac.cn}
\affil{Key Laboratory for Particle Astrophysics, Institute of High Energy Physics, Chinese Academy of Sciences, 19B Yuquan Road, Beijing 100049, China}
\affil{University of Chinese Academy of Sciences, Chinese Academy of Sciences, Beijing 100049, China}

\author{Shu Zhang\textsuperscript{*}}
\email{szhang@ihep.ac.cn}
\affil{Key Laboratory for Particle Astrophysics, Institute of High Energy Physics, Chinese Academy of Sciences, 19B Yuquan Road, Beijing 100049, China}

\author{Long Ji\textsuperscript{*}}
\email{jilong@mail.sysu.edu.cn}
\affil{School of Physics and Astronomy, Sun Yat-Sen University, Zhuhai, 519082, China}

\author{Victor Doroshenko}
\affil{Institut f{\"u}r Astronomie und Astrophysik, Kepler Center for Astro and Particle Physics, Eberhard Karls, Universit{\"a}t, Sand 1, D-72076 T{\"u}bingen, Germany}
\affil{Space Research Institute of the Russian Academy of Sciences, Profsoyuznaya Str. 84/32, Moscow 117997, Russia}

\author{Andrea Santangelo}
\affil{Institut f{\"u}r Astronomie und Astrophysik, Kepler Center for Astro and Particle Physics, Eberhard Karls, Universit{\"a}t, Sand 1, D-72076 T{\"u}bingen, Germany}

\author[0000-0003-0946-3151]{Mauro Orlandini}
\affil{INAF/Osservatorio di Astrofisica e Scienza dello Spazio, Via Gobetti 101, 40129 Bologna, Italy}

\author[0000-0003-2284-571X]{Filippo Frontera}
\affil{Department of Physics and Earth Science, University of Ferrara, Via Saragat, 1 44122 Ferrara, Italy}

\author{Jian Li}
\affil{CAS key Laboratory for Research in Galaxies and Cosmology, Department of Astronomy, Unicersity of Science and Technology of China, Hefei 230026, People's Republic of China}

\author{Yu-Peng Chen}
\affil{Key Laboratory for Particle Astrophysics, Institute of High Energy Physics, Chinese Academy of Sciences, 19B Yuquan Road, Beijing 100049, China}

\author{Peng-Ju Wang}
\affil{Key Laboratory for Particle Astrophysics, Institute of High Energy Physics, Chinese Academy of Sciences, 19B Yuquan Road, Beijing 100049, China}
\affil{University of Chinese Academy of Sciences, Chinese Academy of Sciences, Beijing 100049, China}

\author[0000-0003-4856-2275]{Zhi Chang}
\affil{Key Laboratory for Particle Astrophysics, Institute of High Energy Physics, Chinese Academy of Sciences, 19B Yuquan Road, Beijing 100049, China}

\author{Jin-Lu Qu}
\affil{Key Laboratory for Particle Astrophysics, Institute of High Energy Physics, Chinese Academy of Sciences, 19B Yuquan Road, Beijing 100049, China}

\author{Shuang-Nan Zhang}
\affil{Key Laboratory for Particle Astrophysics, Institute of High Energy Physics, Chinese Academy of Sciences, 19B Yuquan Road, Beijing 100049, China}
\affil{University of Chinese Academy of Sciences, Chinese Academy of Sciences, Beijing 100049, China}


\begin{abstract}
We have performed phase-resolved spectral analysis of the accreting pulsar 1A~0535+262 based on observations of Insight-HXMT during the 2020 Type-\uppercase\expandafter{\romannumeral2} outburst of the source. 
We focus on the two-dimensional dependence of the cyclotron resonance scattering features (CRSFs) along the outburst time and at different phases.
The fundamental CRSF line (f-CRSF) shows different time- and phase-dependent behaviors.
At higher luminosity, the phase profile of the f-CRSF energy changes from a single peak to double peaks, with the transition occurring at MJD 59185. 
On the contrary, the first harmonic CRSF (1-st CRSF) at $\sim$ 100 keV is only detected within a narrow phase range (0.8$-$1.0) accompanied by a shallow f-CRSF line. 
Based on these results, we speculate that when the source enters the supercritical regime, the higher accretion column can significantly enhance the harmonic line at a narrow phase through an "anti-pencil'' beam at a higher energy band. At the same time, it will also affect the behavior of the fundamental line.
 
\end{abstract}

\section{Introduction}
1A~0535+262 is a transient high mass X-ray binary (HMXB) located 2 kpc away (\citealp{Bailer-Jones2018}) with a highly magnetized neutron star (NS) spinning $\sim$ 104 s and orbiting an O9.7IIIe donor star (\citealp{Rosenberg1975, Steele1998}). 
The orbit has an eccentricity of e = $0.47\pm0.02$ and an orbital period of $P_{\rm orb}$ $\sim$ $110.3\pm0.3$ days (\citealp{Finger1996}).

The magnetic field strength $\sim 4\times10^{12}$ G of 1A~0535+262 was measured through the discovery of a fundamental cyclotron resonance scattering feature (CRSF) at $\sim$ 45 keV (\citealp{Kendziorra1994}).
The CRSF energy is related to the magnetic field strength of the NS surface by the ``12-B-12 rule'', $E_{\rm cyc}$ $\approx$ 12$B_{12}$ keV, with the B-field strength $B_{12}$ given in units of $10^{12}$ G (\citealp{Canuto1977}). 
 
A cyclotron line at energy higher than 90 keV was first predicted by \cite{DalFiume88}, who observed the source with a balloon experiment during the 1980 giant outburst, from the comparison between the observed spectrum and the theoretical spectra expected for different values of the magnetic field by \cite{Harding84}.
The presence of a first harmonic line at 110 keV was later observed with \emph{CGRO} OSSE during an outburst in 1994, but the 45 keV feature was not detected (\citealp{Grove1995}).  
During the 2005 type-I outburst of the source, two absorption features were detected in the phase averaged spectra at $\sim45$ keV and $\sim100$ keV (\citealp{Caballero2007}), based on INTEGRAL observations. 
\cite{Caballero2008} also detected a CRSF a slightly higher fundamental energy at $\sim 50$ keV during a pre-outburst flare about five days prior the normal outburst. The authors suggested that this was related to a short accretion episode triggered by magnetospheric instabilities (\citealp{Postnov2008}).

The evolution of the CRSF's parameters with luminosity provides essential information to studying accreting flow geometry and physics near the neutron star surface (\citealp{Maitra2017JApA}). \cite{Caballero2007} considered all CRSF measurements for 1A~0535+262 using data from RXTE, INTEGRAL, Suzaku, and TTM \& HEXE, and found no correlation between the CRSF parameters and luminosity, therefore suggesting that the line forming region does not significantly vary with luminosity. 
Positive $E_{\rm cyc}/L{\rm x}$ correlation was found using “pulse-to-pulse” analysis (\citealp{Klochkov2011}). At the higher luminosities, \cite{Sartore2015} suggested that 1A~0535+26 accretes in the sub-critical regime, and they found no significant variations of the energy of its cyclotron lines with flux by phase-average spectral analysis.
During the brightest 2020 Type-\uppercase\expandafter{\romannumeral2} outburst, \cite{Kong2021}, based on Insight-HXMT data, found a clear transition between positive and negative correlation at a critical luminosity $6.7\times10^{37}$ erg s$^{-1}$, showing that the source changed the accretion regime from sub-critical to super-critical (\citealp{Basko1976, Becker2012, Mushtukov2015MNRAS.447.1847M}). 
Below the critical luminosity, 1A~0535+262 showed an asymmetrical evolution during the outburst, and the CRSF energy kept stable during the decreasing phase, implying the complexity again in understanding the CRSF (\citealp{Kong2021}). 
This finding made 1A0535+262 the second source showing a transition between super- and sub-critical accretion after V0332+53 (\citealp{Doroshenko2017, Vybornov2018}) but with an opposite trend.

\cite{Kong2021} also found that the CRSF centroid energy ratio of the first harmonic line to the fundamental line was $\sim$ 2.3, larger than the factor of 2 predicted from the spacing of the Landau levels. 
Similar results were found for Vela X-1 (\citealp{Kreykenbohm2002}). 
Such behavior can be explained by photon spawning from scattering at higher harmonics (producing lower harmonic photons from a higher harmonic photon). Hence, the depth and shape of the fundamental line can be influenced (\citealp{Schoenherr2007}).
However, \cite{Nishimura2011} explained this behavior as due to a superposition of a large number of line energies that arose from different heights of the accretion column. 

In this work, we present an investigation of the variation of key spectral parameters as a function of the pulse phase, obtained with the finest ever phase and time sampling in a broad energy band. 
Section 2 describes observations and data reduction, and our results are presented in Section 3. 
In Section 4, we discuss our results. 
The summary and conclusions are presented in Section 5.

\section{Observation and Data reduction}

The Hard X-ray Modulation Telescope named Insight-HXMT (\citealp{2014SPIE.9144E..21Z}; \citealp{2020SCPMA..63x9502Z}) was launched on June 15, 2017. The scientific payload includes three collimated telescopes that allow observations in a broad energy band (1-250 keV) and with large effective area at high energies: the High Energy X-ray Telescope (HE, 18 cylindrical NaI(Tl)/CsI(Na) phoswich detectors, \citealp{2020SCPMA..63x9503L}); the Medium Energy X-ray Telescope (ME, 1728 Si-PIN detectors, \citealp{2020SCPMA..63x9504C}); and the Low Energy X-ray Telescope (LE, Swept Charge Device (SCD), \citealp{2020SCPMA..63x9505C}), with collecting-area/energy-range of 5000 $\rm cm^2$/20-250 keV, 952 $\rm cm^2$/5-30 keV and 384 $\rm cm^2$/1-10 keV. The Fields of View (FoVs) are $1.6^{\circ}\times6^{\circ}$, $1^{\circ}\times4^{\circ}$, and $1.1^{\circ}\times5.7^{\circ}$; $5.7^{\circ}\times5.7^{\circ}$ for LE, ME and HE, respectively, with two FOVs for HE (narrow FOV for 15 detection units, large FOV for 2 units, 1 unit with blocked FOV, see Figure 1 in \citealp{Kong2021}). 

Insight-HXMT observed 1A~0535+262 from Nov 6, 2020 (MJD 59159) to Dec 24, 2020 (MJD 59207), for a total exposure of $\sim$ 1.910\,Ms.
In our paper, we select observations at the peak of the outburst on Nov 18, 2020 (MJD 59171) with a long exposure time (LE: 12.14 ks, ME: 12.11 ks, HE: 13.96 ks) as a demonstration for phase-resolved analysis in ten phases ($\sim 1$ ks exposure time for each detector), which can show high signal-to-noise ratio (S/N) of the spectrum above 100 keV and constrain the CRSF energy well.
For the analysis, we use the Insight-HXMT Data Analysis Software (HXMTDAS) v2.04, together with the current calibration model v2.05 (\url{http://hxmtweb.ihep.ac.cn/software.jhtml}). 
The data are selected as recommended by the Insight-HXMT team. In particular, data with elevation angle (ELV) larger than $10^{\circ}$, geometric cutoff rigidity (COR) larger than 8 GeV, and offset for the point position smaller than $0.04^{\circ}$ are used.
In addition, data taken within 300\,s of the South Atlantic Anomaly (SAA) passage outside of good time intervals identified by onboard software have been rejected. 
Contamination from Crab (ra=83.63308, dec=22.0145) has been taken into account as in \cite{Kong2021}.

Based on the results of in-flight calibration (\citealp{Li2020JHEAp}), the energy bands considered for spectral analysis are: $1.5-10$\,keV for the LE, $8-35$\,keV for the ME, and $28-120$\,keV for the HE. 
Because of the uncertain calibration of Ag emission line at $\sim22$ keV of the ME, we ignore the $20-23$ keV range during the spectral analysis for this instrument.
The instrumental backgrounds are estimated with the tools provided by Insight-HXMT team: LEBKGMAP, MEBKGMAP and HEBKGMAP, version 2.0.9 based on the standard Insight-HXMT background models (\citealp{2020arXiv200401432L}; \citealp{2020arXiv200501661L}; \citealp{2020arXiv200306260G}). 
The XSPEC v12.12.0 software package (\citealp{1996ASPC..101...17A}) was used to perform the spectral fitting. To improve the counting statistic of the energy spectra, we combined the exposures within one day by \emph{addspec} and \emph{addrmf} tasks.
Considering the current accuracy of the instrument calibration, we include 0.5$\%$, 0.5$\%$, and 1\% systematic error for spectral analysis for LE, ME, and HE, respectively.
The uncertainties of the spectral parameters are computed using Markov Chain Monte Carlo (MCMC) with a length of 10,000 and are reported at a 90\% confidence level.

\section{Results}

\subsection{Pulse profiles and Phase-resolved spectra}

In the following timing analysis, barycentric and binary orbiting corrections have been performed for period determination and studying of pulse profiles, and the parameters of the binary orbit ($T_0=53613$, $P_{\rm orbit}=111.1$ day, $e=0.47$, $ax\sin i=267$ light-sec, $\omega=130^{\circ}$) are taken from \cite{Finger1996}.
Because of high statistics, the HE observations are used to measure periods in each interval.
To ensure that the phases from different observations are consistent, all HE pulse profiles ($25-80$ keV) with two peaks are aligned to get TOAs through the FFTFIT routine (\citealp{Taylor1992}). We use tempo2 to fit the TOAs evolution (\citealp{Hobbs2006MNRAS}), and get the frequency parameters \footnote{pepoch=59170; $F0=0.0096595755935471$, $F1=1.72126260544263e-11$, $F2=1.11786863370198e-17$, $F3=-5.51116187149374e-23$, $F4=.06922105290258e-28$, $F5=-1.27856062979819e-34$, $F6=7.33508948203979e-41$ for TOA fitting}.

\begin{figure}
    \centering\includegraphics[width=0.49\textwidth,height=23cm]{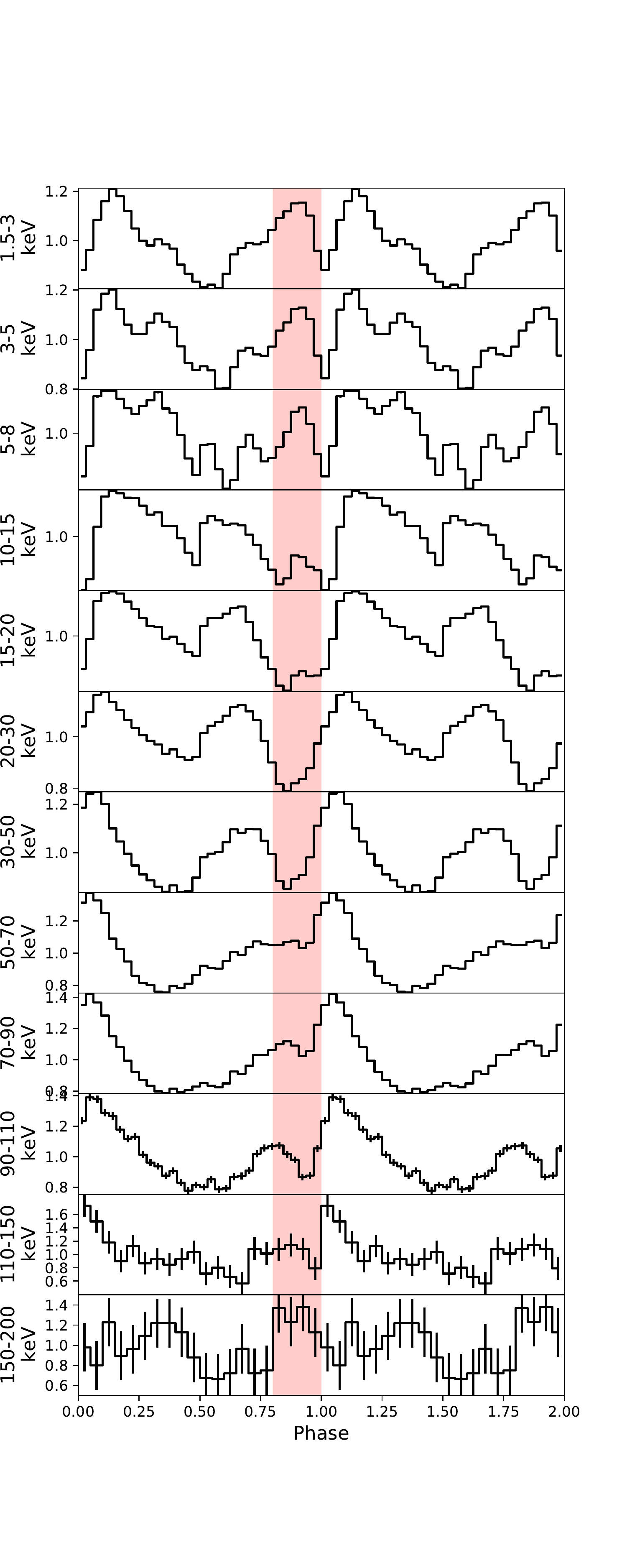}
    \centering\includegraphics[width=0.49\textwidth, height=23cm]{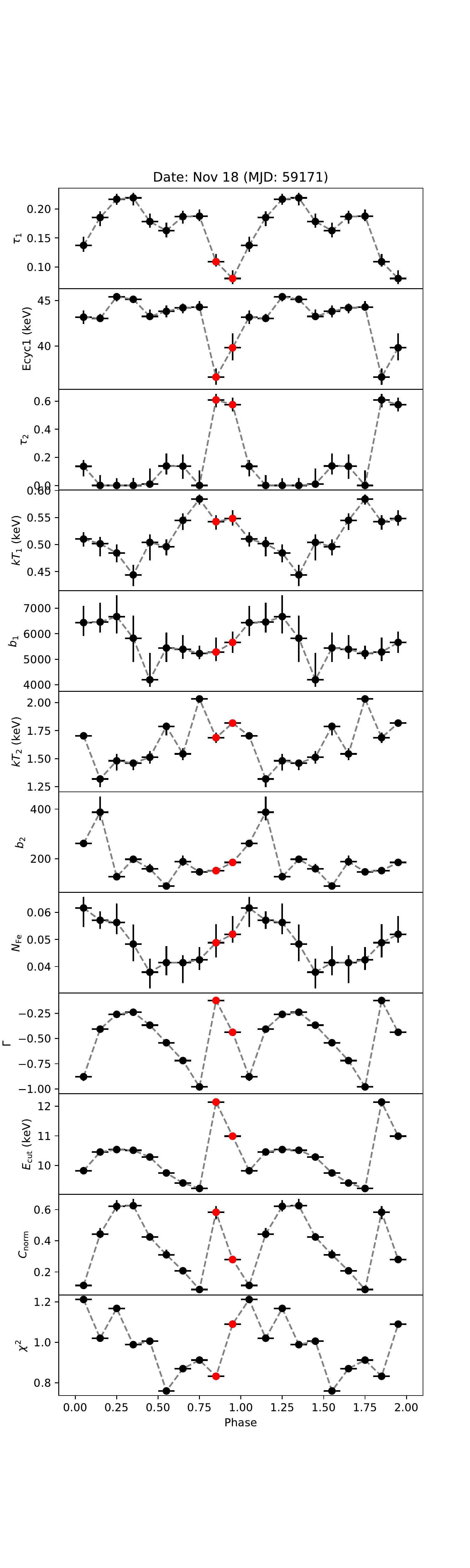}
    \caption{Left panel: The pulse profiles on MJD 59171 in different energy bands. Right panel: The phase-dependent parameters from phase-resolved spectral fittings on MJD 59171. The red points mark the phases with weak fundamental lines and strong 1-st harmonic lines, and the phase range between 0.8 and 1.0 is filled with red color in the left panel.}
    \label{fittings}
\end{figure}

The background-subtracted light curves of LE, ME and HE are folded with twelve energy bands: $1.5-3$ keV, $3-5$ keV, $5-8$ keV, $10-15$ keV, $15-20$ keV, $20-30$ keV, $30-50$ keV, $50-70$ keV, $70-90$ keV, $90-110$ keV, $110-150$ keV, $150-200$ keV. The pulse profiles are normalized by average counts rate.
Taking an observation (Nov 18, 2020) during the peak of the outburst, where the luminosity exceeded the critical one, as an example, the pulse profiles (left panels of Figure~\ref{fittings}) show clear energy dependence: multiple peaks at lower energies ($<10$ keV) and double peaks at higher energies ($>20$ keV). 
And we find that the peaks of pulse profiles simplify at higher energy bands, and a significant dip arises at phase $0.8-1.0$ at $90-110$ keV accompanied by a significant harmonic line at $\sim 100$ keV.
As the energy increases, the profiles change from peak to valley at phase $0.8-1.0$, and the phase separation between the two peaks above 20 keV gradually increases. 
The behaviors of the pulse profiles' evolution with energy might be related to different relativistic beaming angles at various energy bands and will be discussed in Sec 4.

\begin{figure}
    \centering\includegraphics[angle=-90,width=0.49\textwidth]{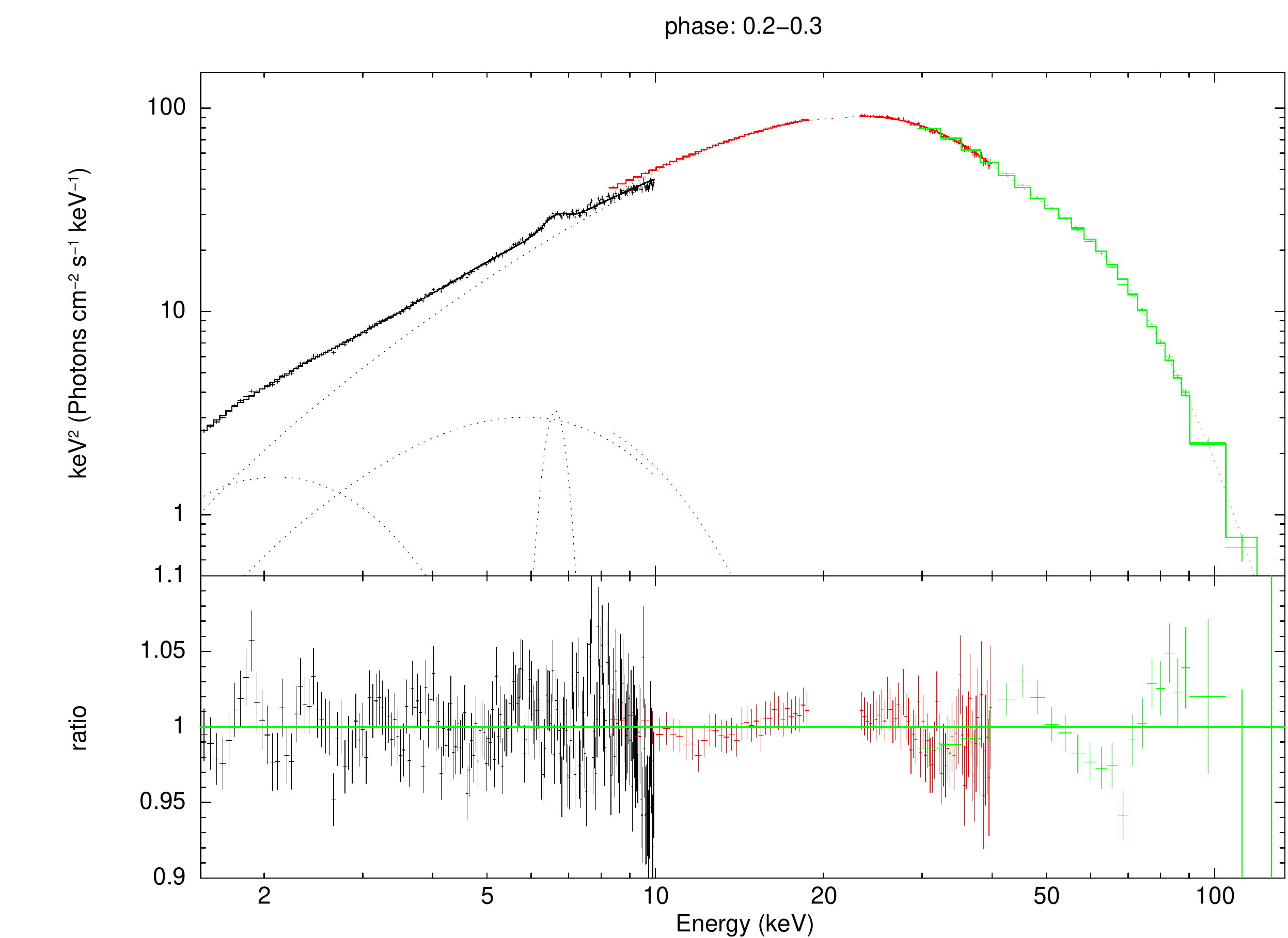}
    \centering\includegraphics[angle=-90,width=0.49\textwidth]{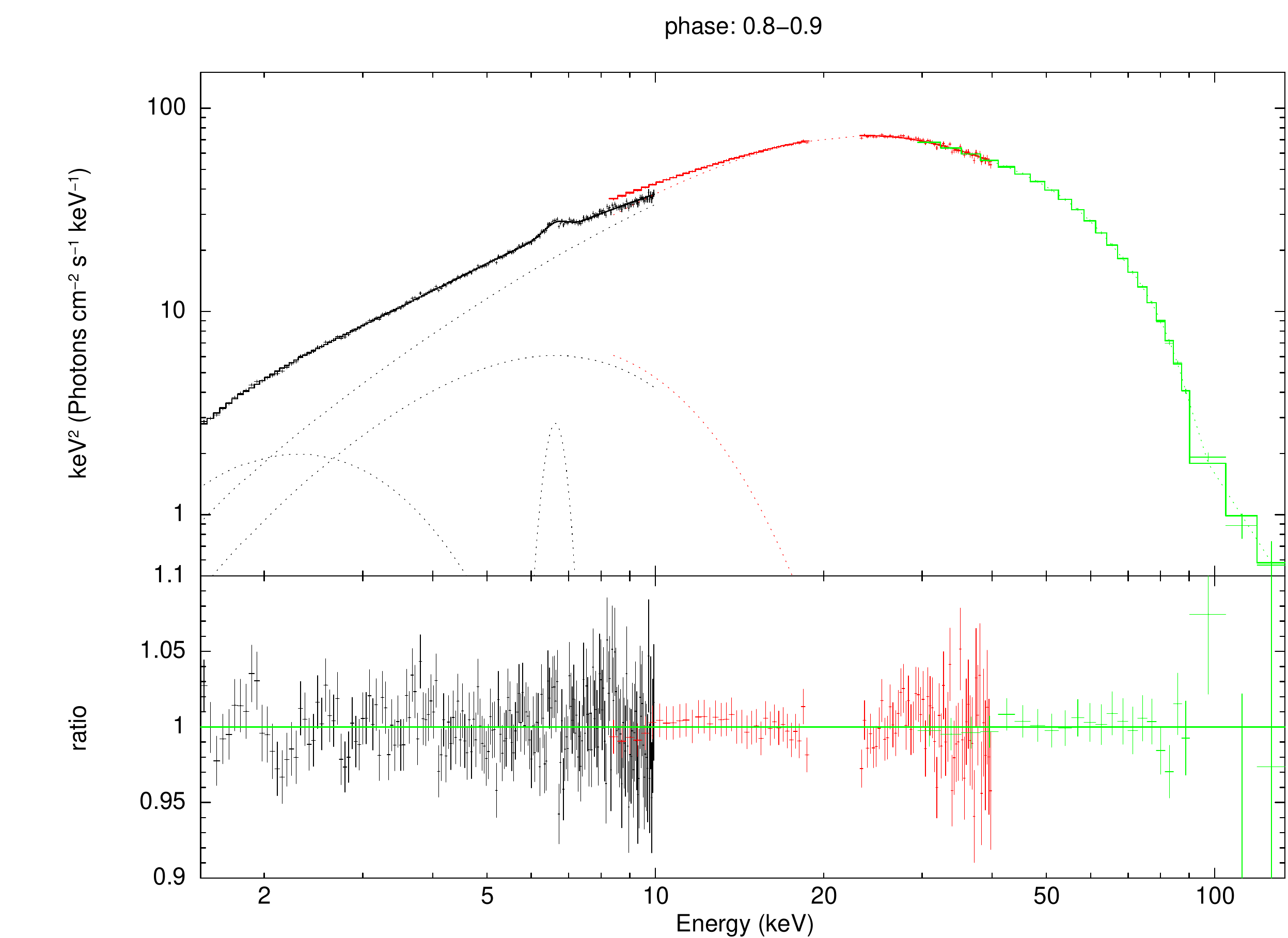}
    \centering\includegraphics[angle=-90,width=0.49\textwidth]{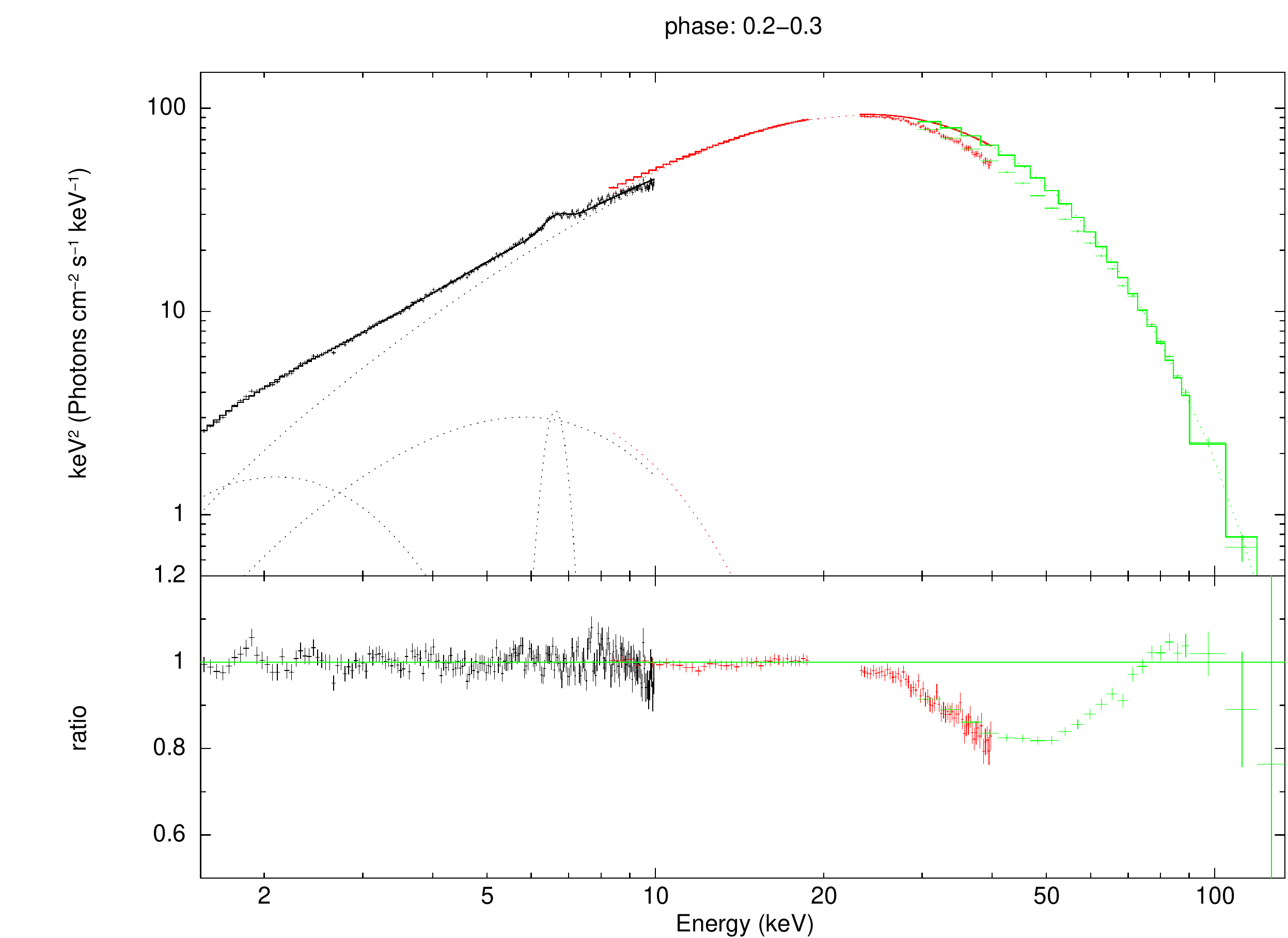}
    \centering\includegraphics[angle=-90,width=0.49\textwidth]{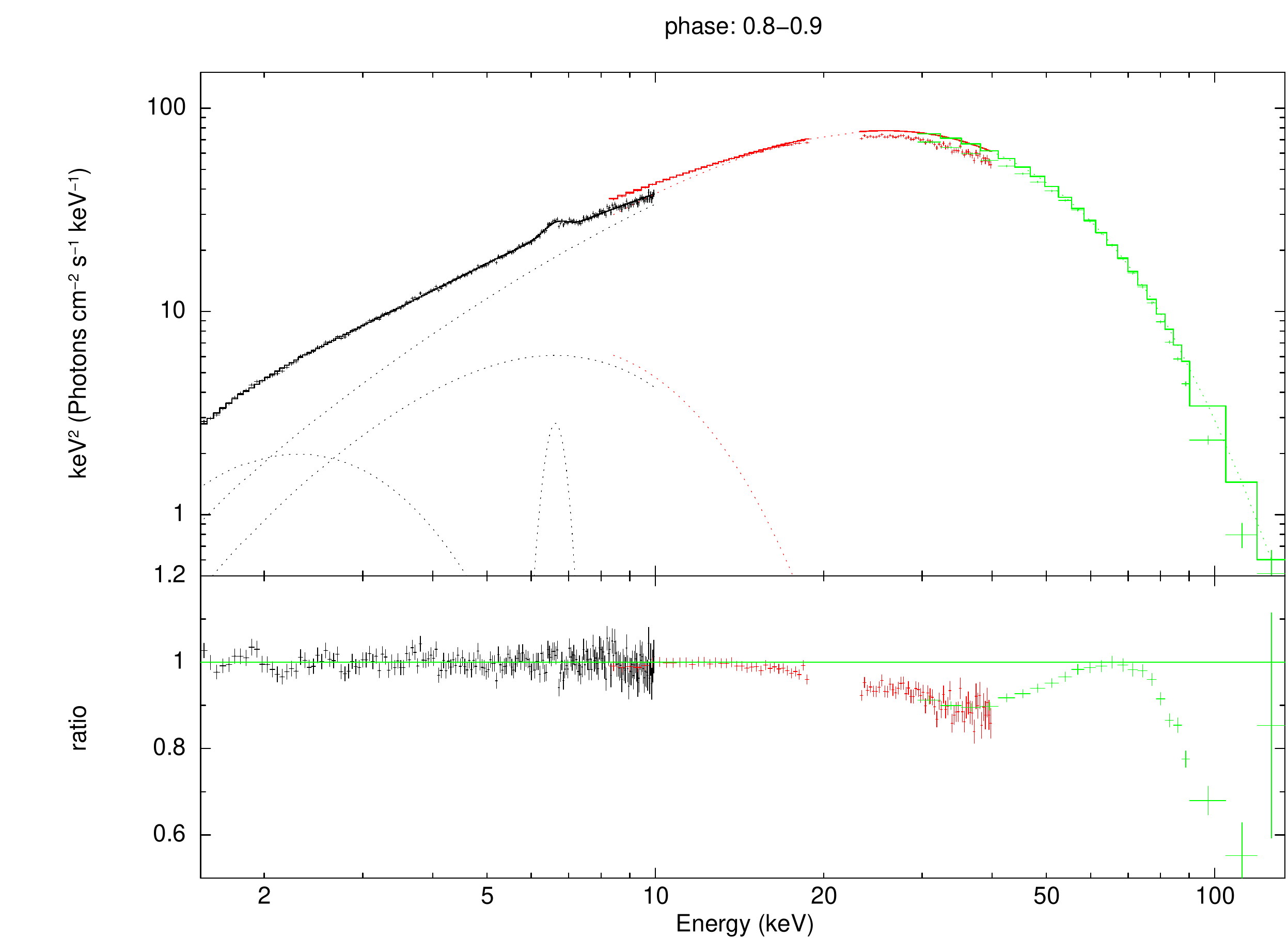}
    \caption{The spectral fittings and residuals for phase 0.2-0.3 (left) and 0.8-0.9 (right) on MJD 59171. The lower panels of the two phases show the spectral fittings and residuals without the CRSF lines.}
    \label{spectra}
\end{figure}

In Figure~\ref{fittings} (right panel), Figure~\ref{spectra} and Table~\ref{spectral_fitting}, we show the observation on Nov 18, 2020 (combine ObsID P0304099007, P0314316002, P0314316003 within one day) as an example.
For each observation, we select the photons within ten different phases, and the spectra are extracted from these decadal phase intervals. We use the same model \emph{Tbabs$\times$mgabs$\times$(bbodyrad1+bbodyrad2+cutoffpl+gaussian)} in \cite{Kong2021} to fit these spectra at different phases. 
The \emph{cutoffpl} is a simple continuum with just three free parameters:
\begin{equation}
F(E)=K \times E^{-\Gamma}exp(-E/E_{\rm fold}), 
\end{equation}
where $K$, $\Gamma$ and $E_{\rm fold}$ determine the normalization coefficient, the photon index, and the exponential folding energy, respectively.
The residuals at lower energies are accounted for by adding two black-body components: a cooler one with $kT<1$ keV (\emph{bbodyrad1}) and a hotter one with $kT>1$ keV (\emph{bbodyrad2}). 
The ISM absorption is taken into account through the \emph{tbabs} model (\citealp{Wilms2000}), where $n_{\rm H}$, is fixed at $0.59 \times 10^{22}$ atom cm$^{-2}$. 
A \emph{gaussian} line is needed to model the iron emission line, and we fix the energy at 6.6 keV and the width at 0.3 keV.
To model the CRSF absorption features at $\sim45$ and $\sim 100$\,keV which are visible in the phase-average spectra in \cite{Kong2021}, we adopt a multiplicative absorption model \emph{mgabs} with Gaussian profile:
\begin{equation}
F^{'}(E) = F(E) \times \emph{mgabs} = F(E) \times [(1-\tau_{1}e^{\frac{-(E-E_{\rm cyc1})^2}{2\sigma_{1}^{2}}}) \times (1-\tau_{2}e^{\frac{-(E-E_{\rm cyc2})^2}{2\sigma_{2}^{2}}})], 
\end{equation}
where $F^{'}(E)$ is the spectrum modified by \emph{mgabs}, $E_{\rm cyc1}$ is the cyclotron line central energy of the fundamental line, $\tau_{1}$ and $\sigma_{1}$ characterize the central absorption depth and the width of the line. 
Same as fundamental line, $E_{\rm cyc2}$, $\tau_{2}$ and $\sigma_{2}$ describe the central energy, the depth, and the width of the 1-st harmonic line.
The energy and width of the 1-st harmonic line are fixed at 100 keV and 10 keV during the spectral fitting because of the high background above 100 keV and relatively low statistics. 
We note that the width of the 100 keV line was fixed at 5 keV in \cite{Kong2021}, but for phase-resolved spectra at phase 0.8-1.0, it leaves larger $\chi^2$ and residuals. 
The reported uncertainties for the best-fit parameters are estimated using Markov Chain Monte Carlo (MCMC) method with a chain length of 10,000.
The parameters and reduced $\chi^2$ are plotted in the right panel in Figure~\ref{fittings} and listed in Table~\ref{spectral_fitting}.
From the spectral fitting, for the fundamental line we find that the $E_{\rm cyc1}$ is essentially constant with a value of $\sim44$ keV except for phases $0.8-0.9$ where it drops to $\sim37$ keV.
The absorption depth $\tau_1$ stays stable $\sim 0.15$ except at $0.8-1.0$ with a dramatic drop to $\sim0.08$.
The phase dependence of the depth of the fundamental line can be described as showing two peaks at pulse phases of $0.2-0.3$ and $0.7-0.8$, and a dip at phase $0.8-0.9$.
The depth, $\tau_2$, of the first harmonic shows a significant increase at the same phase range, whereas $\tau_1$ exhibits a considerable decrease.
$\Gamma$ and $E_{\rm cut}$ of \emph{cutoffpl} have abrupt increases at phase $0.8-1.0$ and they show a correlated behavior versus phase.
In Figure~\ref{spectra}, we show the best-fit fits for spectra and corresponding residuals with and without the fundamental and the first harmonic CSRF lines at phase $0.2-0.3$ and $0.8-0.9$, respectively. 

\begin{figure}
    \centering\includegraphics[width=0.49\textwidth]{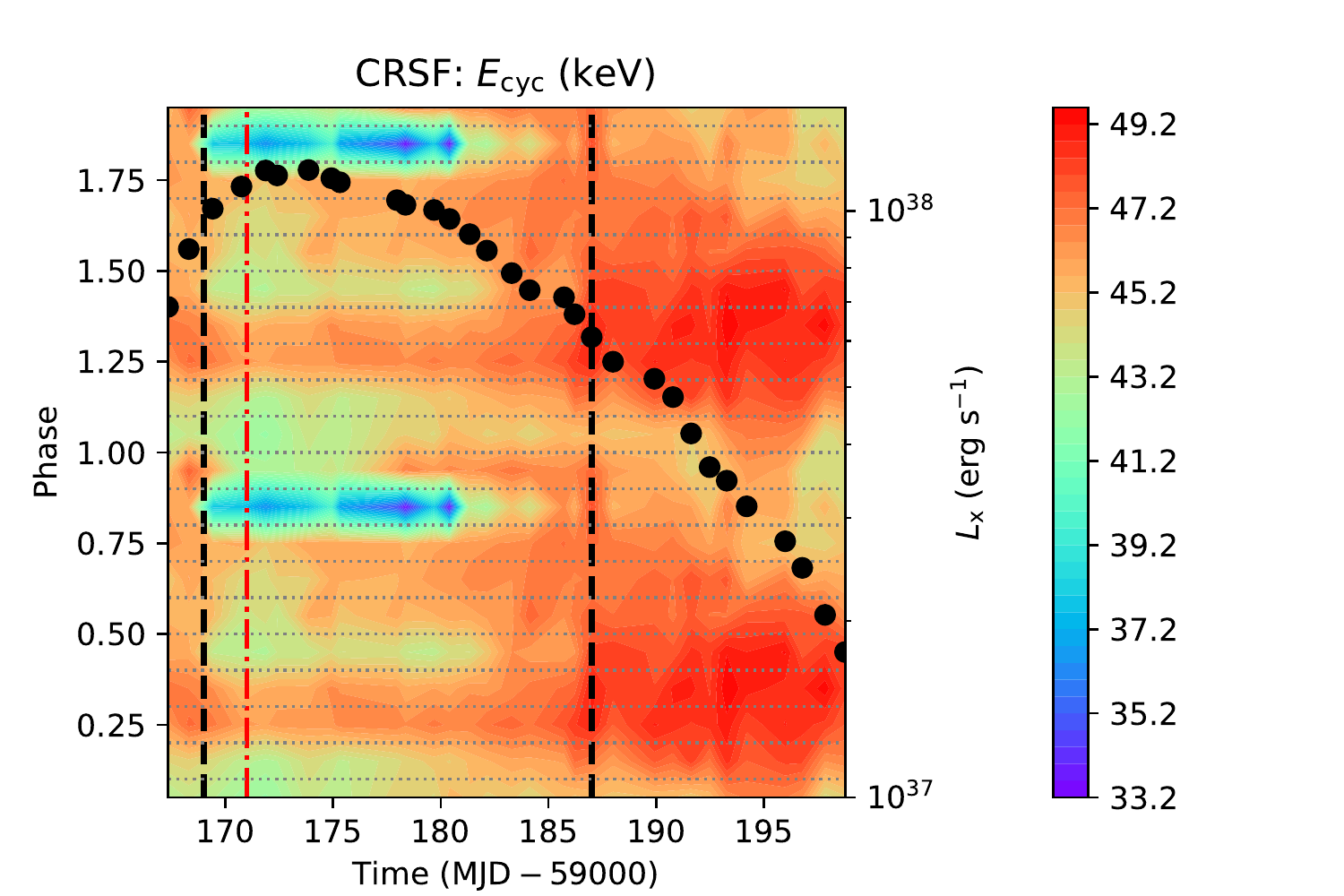}
    \centering\includegraphics[width=0.49\textwidth]{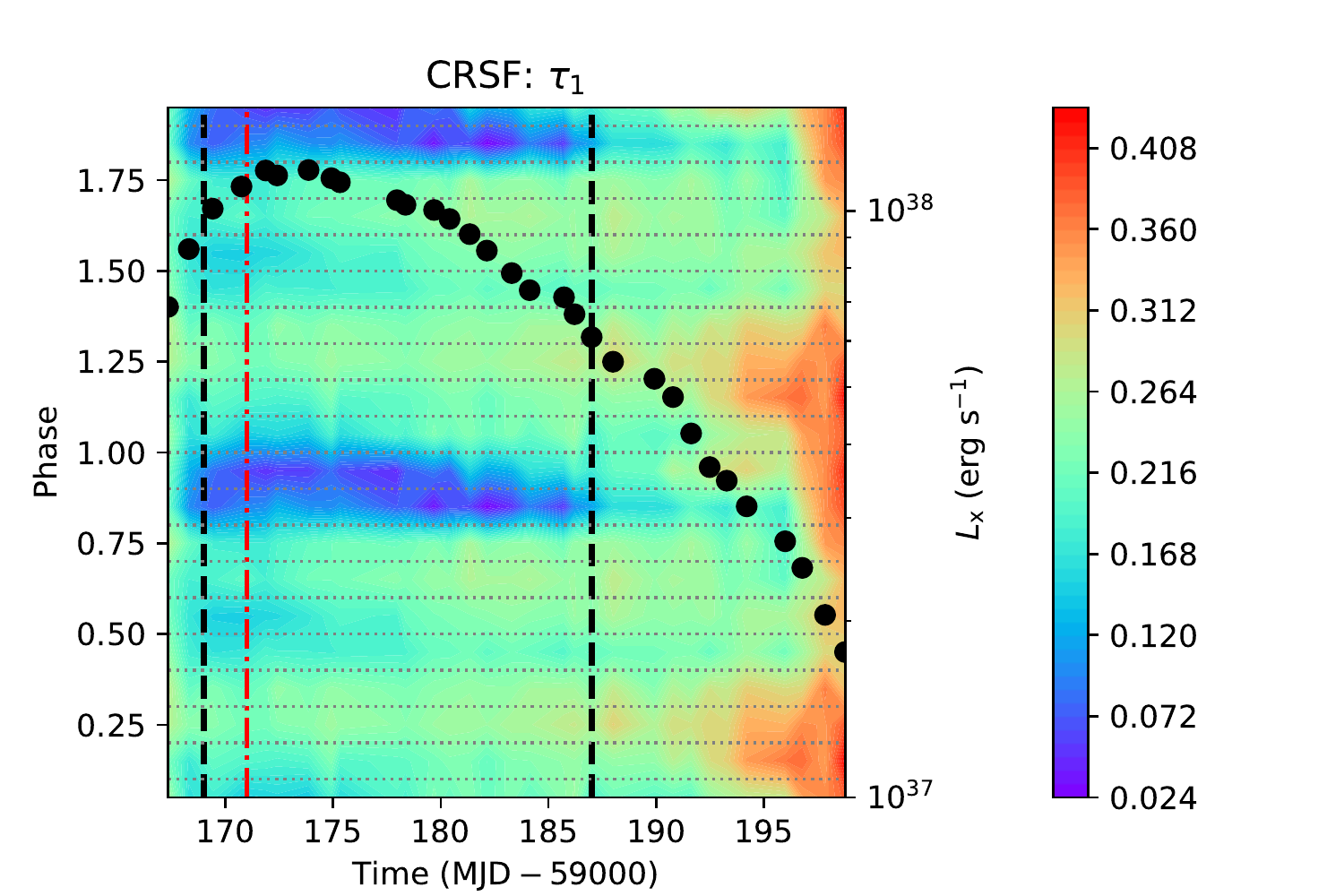}
    \\
    \centering\includegraphics[width=0.49\textwidth]{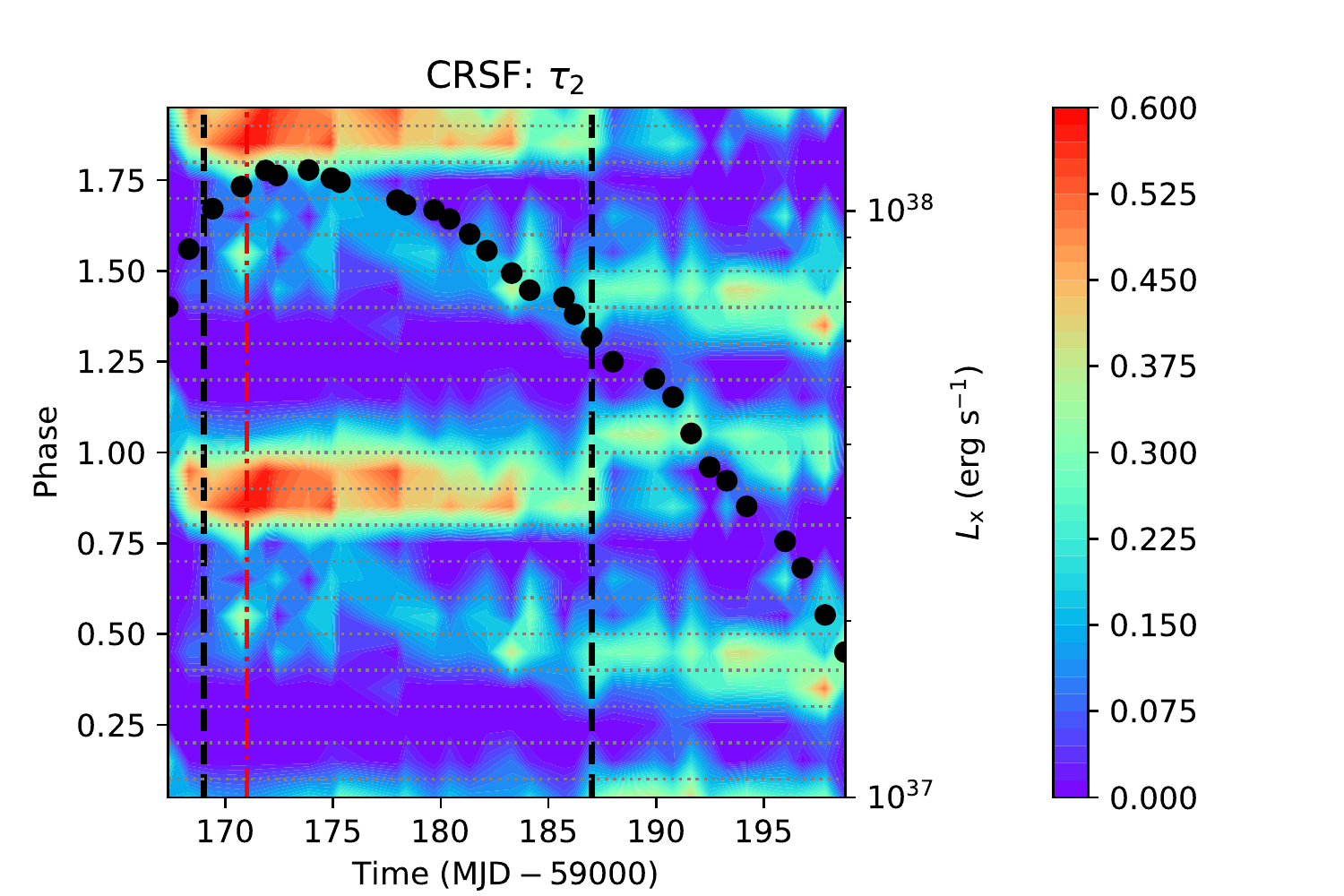}
    \centering\includegraphics[width=0.49\textwidth]{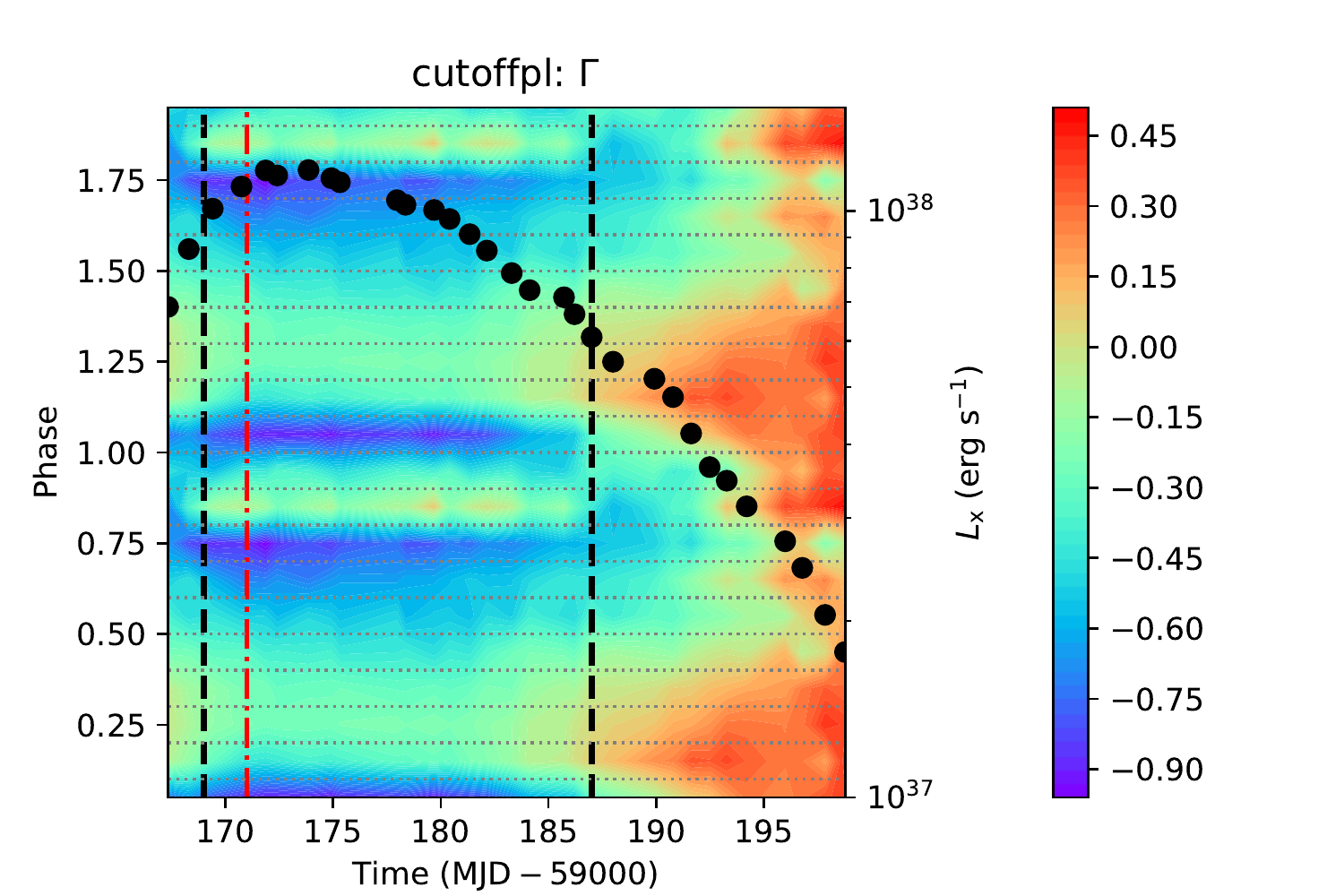}
    \\
    \centering\includegraphics[width=0.49\textwidth]{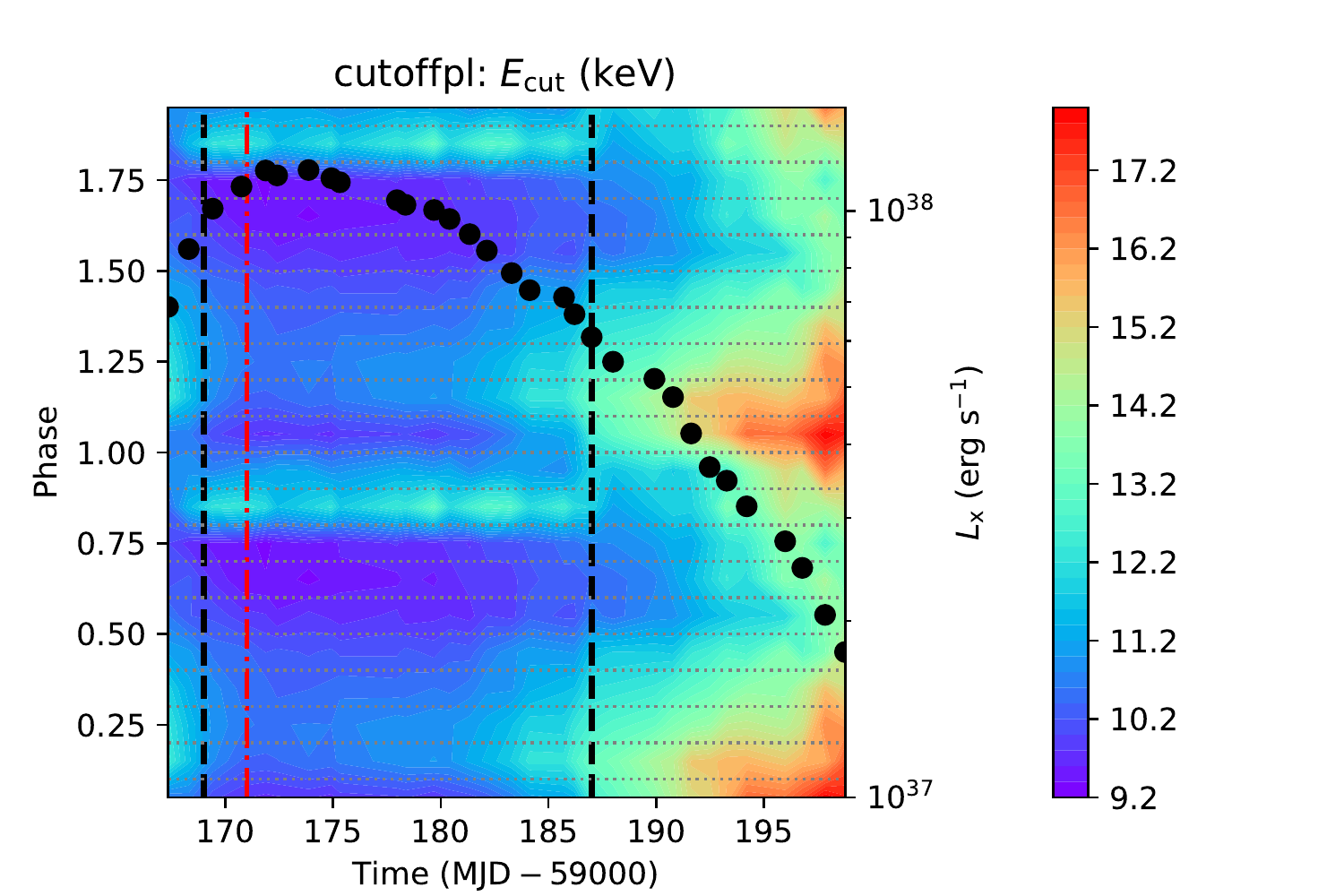}
    \centering\includegraphics[width=0.49\textwidth]{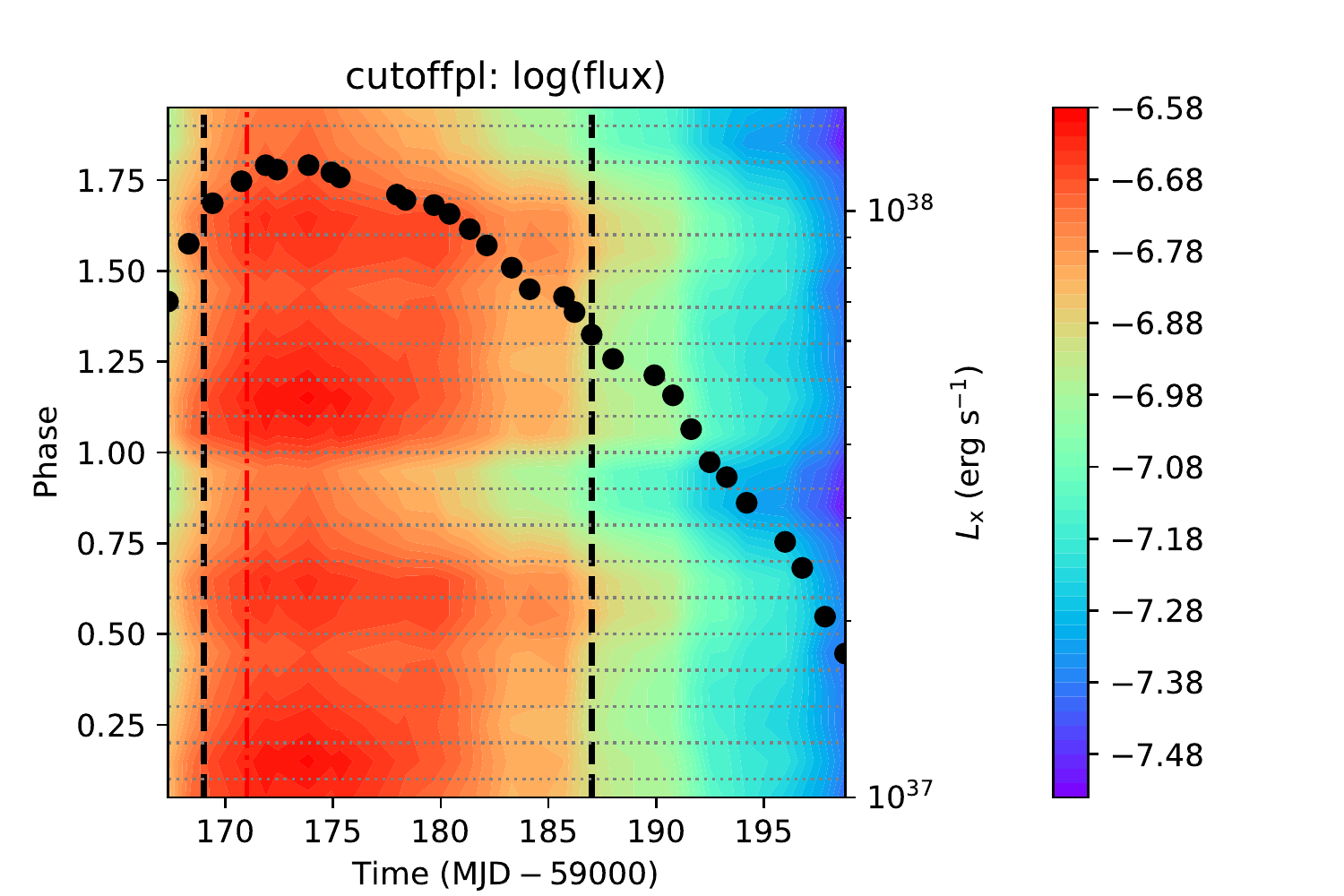}
    \caption{The 2-D MJD vs Phases distributions (2-D MPDs) show the CRSF energy $E_{\rm cyc}$, the absorption depth $\tau_{1}$ at fundamental, absorption depth at first harmonic $\tau_{2}$; photon index $\Gamma$, high-energy cut $E_{\rm cut}$ and $\log$(flux) of \emph{cutoffpl} component in $2-150$ keV. Two black dashed lines mark the time range between MJD 59169 and 59187 which is associated with the luminosity above the critical luminosity.
    The red dashed and dotted lines represent the MJD 59171 at the peak of the outburst.  Gray dotted lines separate ten phases intervals. The black points show how luminosities evolve with time.}
    \label{2-D MPDs}
\end{figure}

\begin{figure}
    \centering\includegraphics[width=0.49\textwidth]{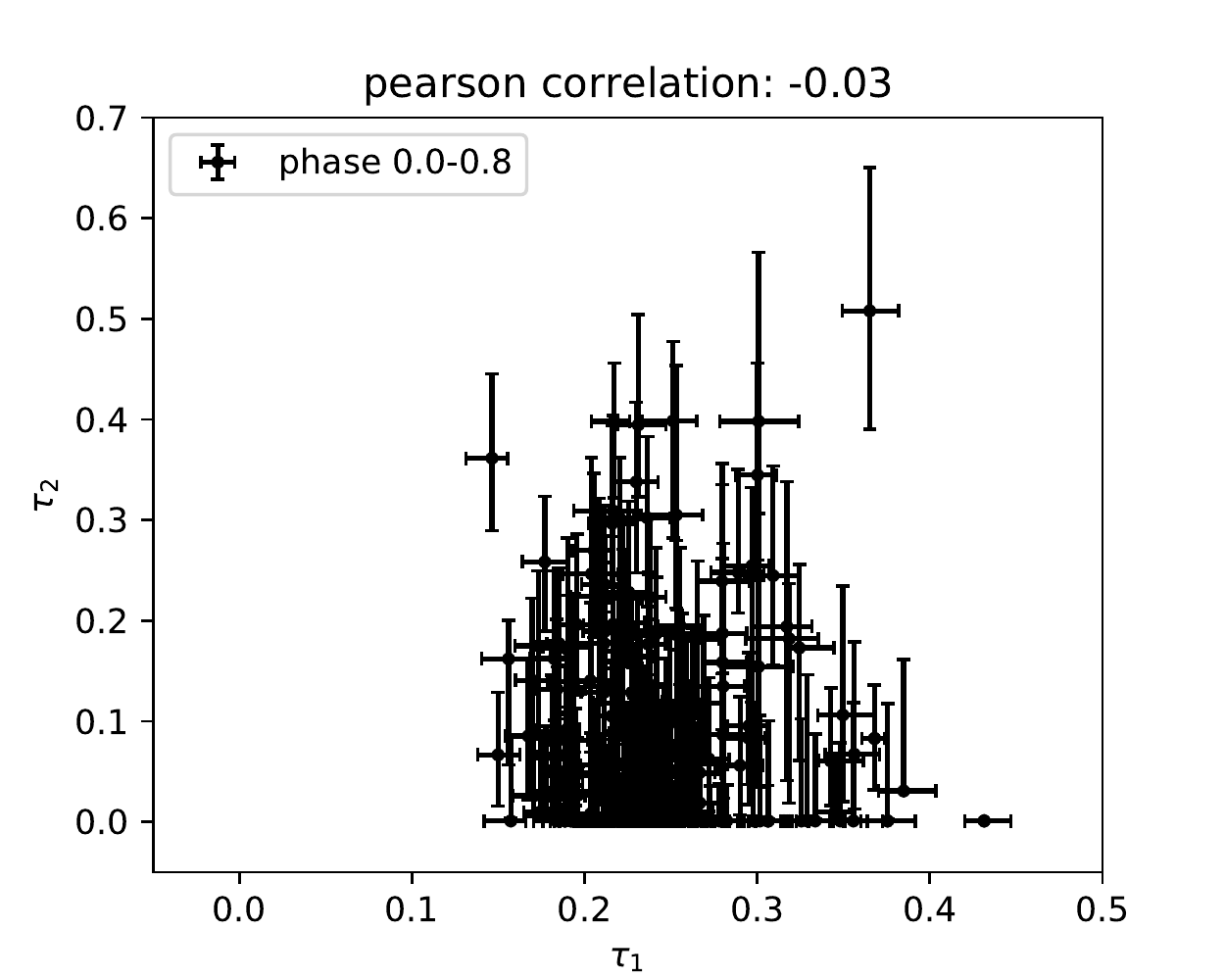}    \centering\includegraphics[width=0.49\textwidth]{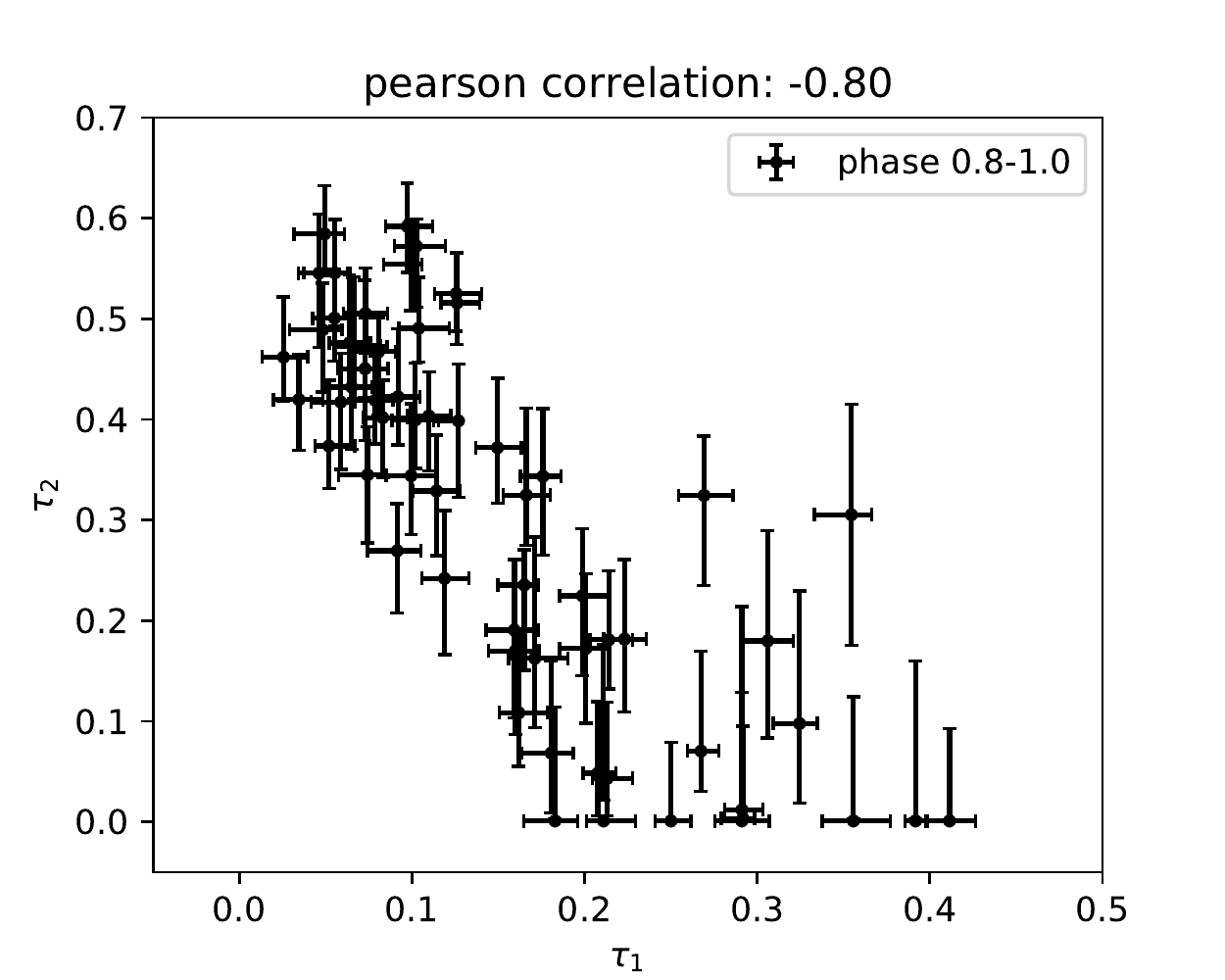}
    \caption{By showing the fitting results of absorption depth in all observations, there is no clear correlation in the left panel (Pearson correlation factor, $\rho=-0.03$) of $\tau_1$ and $\tau_2$ between phase $0.0-0.8$, but a clear anti-correlation is shown in the right panel between $0.8-1.0$ with large $\rho=-0.8$. Above the critical luminosity, most $\tau_2$ between phase $0.0-0.8$ is near zero.}
    \label{cor}
\end{figure}

\subsection{2-D distribution}
The phase dependence and time evolution of the CRSF parameters can also be visualized using two-dimensional color maps (2-D maps). 
Such maps are shown in Figure~\ref{2-D MPDs} for parameters of the CRSF energy and its optical depth for the fundamental and harmonic lines, the photon index $\Gamma$, the high energy cut $E_{\rm cut}$ and the $\log$(flux) of the \emph{cutoffpl} component in $2-150$ keV. Phase dependence of all parameters exhibits an apparent change around MJD 59186.

First, the intensity and the energy $E_{\rm cyc}$ of the fundamental line shows a single broad sinusoidal profile between MJD 59186 and 59198 during the subcritical regime, as demonstrated in the right panel in Figure~\ref{45_distri}. 
They convert to a double-peaks shape with small modulation between MJD 59169 and 59185 during the supercritical regime, as shown in the upper left panel of Figure~\ref{45_distri}, if we ignore the dip between phase $0.8-0.9$.
The CRSF energies between phases $0.0-0.8$ are higher than those between $0.8-0.9$, with a modulation essentially near $\sim45$ keV except a drop to $\sim40$ keV.
From the bottom panels of Figure~\ref{45_distri}, the absorption depth $\tau_1$ keeps double-peak shape within the whole outburst, while in Figure~\ref{2-D MPDs}, the $\tau_2$ shows significant absorption at phase $0.8-1.0$ between MJD 59169 and 59185.
We find that the absorption depths of $\tau_1$ and $\tau_2$ are inversely correlated at phase $0.8-1.0$ (see Figure~\ref{cor}).
The harmonic line is only visible in a narrow phase of $0.8-1.0$, where the absorption depth of fundamental lines become weaker ($\tau_1$ $\leq$ 0.1) and $E_{\rm cyc}$ $\leq$ 40 keV between MJD 59169 and 59180.

Second, a transition of the non-thermal component from a broad single peak to double peaks, accompanied by a narrow ridge in phase $0.8-1.0$, happens when luminosity exceeds $L_{\rm crit}$.
This change is reflected in the Figure~\ref{2-D MPDs} through the evolution of parameters of the photon index $\Gamma$ and cut off energy $E_{\rm cut}$.

\begin{sidewaystable}[ptbptbptb]
\begin{center}
\caption{Parameters of the spectral fitting}
\resizebox{.95\columnwidth}{!}{
\begin{tabular}{cccccccccccccccc}
\hline
\hline
phase & & $0.0-0.1$ & $0.1-0.2$ & $0.2-0.3$ & $0.3-0.4$ & $0.4-0.5$ & $0.5-0.6$ & $0.6-0.7$ & $0.7-0.8$ & $0.8-0.9$ & $0.9-1.0$
\\
\hline
TBabs & $n_{\rm H}\ (10^{22}\ \rm cm^{-2})$ & $0.59$ (fixed) & $0.59$ (fixed) & $0.59$ (fixed) & $0.59$ (fixed) & $0.59$ (fixed) & $0.59$ (fixed) & $0.59$ (fixed) & $0.59$ (fixed) & $0.59$ (fixed) & $0.59$ (fixed)
\\
mgabs & $E_{\rm cyc1}$ (keV) & $43.18_{-0.76}^{+0.73}$ &$43.04_{-0.32}^{+0.50}$ &$45.41_{-0.50}^{+0.37}$ &$45.13_{-0.35}^{+0.37}$ &$43.26_{-0.38}^{+0.74}$ &$43.81_{-0.66}^{+0.65}$ &$44.21_{-0.60}^{+0.45}$ &$44.28_{-0.34}^{+0.66}$ &$36.61_{-0.85}^{+0.95}$ &$39.82_{-1.39}^{+1.58}$
\\
& $\sigma_{1}$ (keV) & $10$ (fixed) & $10$ (fixed) & $10$ (fixed) & $10$ (fixed) & $10$ (fixed) & $10$ (fixed) & $10$ (fixed) & $10$ (fixed) & $10$ (fixed) & $10$ (fixed)
\\
& $\tau_1$ & $0.14_{-0.01}^{+0.01}$ &$0.19_{-0.01}^{+0.01}$ &$0.22_{-0.01}^{+0.01}$ &$0.22_{-0.01}^{+0.01}$ &$0.18_{-0.01}^{+0.01}$ &$0.16_{-0.01}^{+0.01}$ &$0.19_{-0.01}^{+0.01}$ &$0.19_{-0.01}^{+0.01}$ &$0.11_{-0.01}^{+0.01}$ &$0.08_{-0.01}^{+0.01}$
\\
& $E_{\rm cyc2}$ (keV) & $100$ (fixed) & $100$ (fixed) & $100$ (fixed) & $100$ (fixed) & $100$ (fixed) & $100$ (fixed) & $100$ (fixed) & $100$ (fixed) & $100$ (fixed) & $100$ (fixed)
\\
& $\sigma_{2}$ (keV) & $10$ (fixed) & $10$ (fixed) & $10$ (fixed) & $10$ (fixed) & $10$ (fixed) & $10$ (fixed) & $10$ (fixed) & $10$ (fixed) & $10$ (fixed) & $10$ (fixed)
\\
& $\tau_2$ & $0.14_{-0.07}^{+0.05}$ & $\leq0.001$ & $\leq0.001$ & $\leq0.001$ & $\leq0.001$ &$0.14_{-0.06}^{+0.09}$ &$0.14_{-0.09}^{+0.08}$ & $\leq0.001$ &$0.61_{-0.05}^{+0.04}$ &$0.57_{-0.05}^{+0.05}$
\\
gaussian & $E_{\rm Fe}$ (keV) & 6.6 (fixed) & 6.6 (fixed) & 6.6 (fixed) & 6.6 (fixed) & 6.6 (fixed) & 6.6 (fixed) & 6.6 (fixed) & 6.6 (fixed) & 6.6 (fixed) & 6.6 (fixed)
\\
& $\sigma_{\rm Fe}$ (keV) & 0.3 (fixed) & 0.3 (fixed) & 0.3 (fixed) & 0.3 (fixed) & 0.3 (fixed) & 0.3 (fixed) & 0.3 (fixed) & 0.3 (fixed) & 0.3 (fixed) & 0.3 (fixed)
\\
& norm & $0.0617_{-0.0071}^{+0.0041}$ &$0.0571_{-0.0032}^{+0.0032}$ &$0.0563_{-0.0043}^{+0.0070}$ &$0.0483_{-0.0064}^{+0.0073}$ &$0.0379_{-0.0060}^{+0.0050}$ &$0.0414_{-0.0047}^{+0.0061}$ &$0.0414_{-0.0076}^{+0.0028}$ &$0.0425_{-0.0038}^{+0.0047}$ &$0.0488_{-0.0054}^{+0.0068}$ &$0.0519_{-0.0031}^{+0.0068}$
\\
bbodyrad1 & $kT$ (keV) & $0.51_{-0.01}^{+0.01}$ &$0.50_{-0.02}^{+0.01}$ &$0.48_{-0.02}^{+0.02}$ &$0.44_{-0.02}^{+0.02}$ &$0.50_{-0.03}^{+0.01}$ &$0.50_{-0.02}^{+0.01}$ &$0.54_{-0.02}^{+0.01}$ &$0.58_{-0.01}^{+0.01}$ &$0.54_{-0.01}^{+0.01}$ &$0.55_{-0.01}^{+0.02}$
\\
& norm & $6437.91_{-523.48}^{+649.23}$ &$6459.86_{-417.76}^{+752.62}$ &$6672.35_{-662.66}^{+838.79}$ &$5824.91_{-938.32}^{+884.70}$ &$4195.73_{-268.86}^{+1058.65}$ &$5443.19_{-553.27}^{+604.55}$ &$5393.00_{-375.26}^{+551.23}$ &$5227.32_{-227.66}^{+304.30}$ &$5284.12_{-351.90}^{+569.00}$ &$5660.89_{-417.81}^{+429.54}$
\\
bbodyrad2 & $kT$ (keV) & $1.70_{-0.03}^{+0.02}$ &$1.32_{-0.07}^{+0.03}$ &$1.48_{-0.09}^{+0.06}$ &$1.46_{-0.06}^{+0.03}$ &$1.51_{-0.06}^{+0.06}$ &$1.79_{-0.08}^{+0.04}$ &$1.54_{-0.06}^{+0.05}$ &$2.03_{-0.03}^{+0.03}$ &$1.69_{-0.05}^{+0.04}$ &$1.82_{-0.02}^{+0.03}$
\\
& norm & $261.89_{-9.10}^{+12.79}$ &$387.40_{-32.54}^{+63.42}$ &$128.13_{-12.42}^{+15.52}$ &$198.37_{-9.95}^{+15.02}$ &$160.14_{-12.10}^{+19.59}$ &$90.75_{-7.44}^{+9.61}$ &$188.88_{-16.51}^{+24.33}$ &$147.63_{-6.37}^{+5.46}$ &$152.63_{-10.58}^{+12.01}$ &$185.66_{-10.81}^{+10.01}$
\\
cutoffPL & $\Gamma$ & $-0.88_{-0.04}^{+0.02}$ &$-0.41_{-0.02}^{+0.03}$ &$-0.26_{-0.02}^{+0.03}$ &$-0.24_{-0.01}^{+0.03}$ &$-0.37_{-0.02}^{+0.03}$ &$-0.54_{-0.03}^{+0.04}$ &$-0.72_{-0.03}^{+0.03}$ &$-0.98_{-0.02}^{+0.04}$ &$-0.12_{-0.03}^{+0.03}$ &$-0.44_{-0.03}^{+0.03}$
\\
& $E_{\rm fold}$ & $9.82_{-0.11}^{+0.06}$ &$10.45_{-0.08}^{+0.11}$ &$10.54_{-0.06}^{+0.08}$ &$10.51_{-0.03}^{+0.08}$ &$10.28_{-0.05}^{+0.10}$ &$9.75_{-0.06}^{+0.10}$ &$9.41_{-0.09}^{+0.06}$ &$9.22_{-0.04}^{+0.10}$ &$12.14_{-0.14}^{+0.13}$ &$10.99_{-0.10}^{+0.12}$
\\
& norm & $0.11_{-0.01}^{+0.01}$ &$0.44_{-0.02}^{+0.04}$ &$0.62_{-0.03}^{+0.04}$ &$0.62_{-0.02}^{+0.04}$ &$0.42_{-0.02}^{+0.03}$ &$0.31_{-0.02}^{+0.03}$ &$0.21_{-0.01}^{+0.01}$ &$0.09_{-0.01}^{+0.01}$ &$0.58_{-0.04}^{+0.04}$ &$0.28_{-0.02}^{+0.02}$
\\
Fitting & $\chi^{2}_{red}$/d.o.f & 1.21/305 & 1.02/305 & 1.17/305 & 0.99/305 & 1.01/305 & 0.76/305 & 0.87/305 & 0.91/305 & 0.83/305 & 1.09/305 
\\
\hline
\hline
\end{tabular}
\label{spectral_fitting}
}    
\begin{list}{}{}
    \item[Note]{: Uncertainties are reported at the 90\% confidence interval and were computed using MCMC (Markov Chain Monte Carlo) of length 10,000. The 0.5\%, 0.5\%, 1\% system error for LE, ME, and HE have been added during spectral fittings.}
\end{list}
\end{center}
\end{sidewaystable}

\section{Discussion}
Based on the Insight-HXMT's high cadence and high statistic observations which covered the the entire Giant (Type-\uppercase\expandafter{\romannumeral2}) outburst of 1A~0535+262 in 2020, we have performed the detailed phase-resolved spectral analysis upon both the CRSF and continuum spectral components, and investigated their evolution throughout the outburst that covers a luminosity range of $7\times10^{36}-1.2\times10^{38}$ erg s$^{-1}$. 
In \cite{Kong2021}, a significant anti-correlation between the fundamental CRSF energy and luminosities was found for the first time in this source above $6.7\times10^{37}$ erg s$^{-1}$ based on the phase-average analysis, which was associated with a transition from sub- to super-critical accretion regime. 
As shown in Figure~\ref{2-D MPDs} where the two regimes are indicated with the dashed lines, the two regimes are characterized by different intrinsic beam patterns of emerging radiation and, correspondingly, a strong transition of the observed spectrum can be expected. 
In V~0332+53, \cite{Lutovinov2015} found the spectral parameters modulate with phases show different behaviors under different pulse profiles (see Figure 2 in their paper). Still, they did not show the detail of the evolution with time. 
The parameters of non-thermal components and CRSFs changing with the phases are correlated to the optical depth, electron temperature, magnetic field intensity, and cross-section in the accretion column under different viewing angles. For 1A~0535+262, using the high cadence observations and phase-resolved spectral analysis, we can investigate parameters modulated with phases in detail.

In the left panel of Figure~\ref{fittings}, the pulse profiles at luminosity $\sim1.2\times10^{38}$ erg s$^{-1}$ (MJD 59171) are dominated by two peaks above 15 keV, which are generally attributed to the fan beam pattern above the critical luminosity (\citealp{Davidson1973NPhS}). 
However, several peaks or dips underline the complexity of the pulse profile in the lower energy band.
Sec 3.1 noted that the two peaks' evolution with energy might be related to the relativistic beaming effect, especially the phase separation between the two peaks. Understanding this is difficult because of the radiation transfer complexity in the accretion column. For the beaming pattern, the higher the falling velocity, the more radiation beam escaping from the column wall down to the neutron star surface rather than perpendicular to the magnetic field, which makes the phase separation of the two peaks of the fan-beam moving further away. 
Thus, the assumption is that the falling material at the higher position of the accretion column has a larger velocity, leading to the high-energy radiation coming from the higher part of the column.

Above the critical luminosity, the parameters in different phases also show their complexity. 
In the right panel of Figure~\ref{fittings} and panels in Figure~\ref{2-D MPDs}, the parameters of non-thermal component \emph{cutoffpl} show double peaks, including a wide one and a narrow one.
Interestingly, the narrow peak is only present at phase $0.8-1.0$ when the outburst steps into the supercritical area between MJD 59169 and 59187. The sudden rising of $\Gamma$ and $E_{\rm cut}$ imply the spectra become softer but with more high energy photons, which related to the increasing of pulse ratio in $50-70$ keV and $70-90$ keV.
Meanwhile, at the same phase, the narrow peak below 15 keV might contribute to more soft photons and the softening of the continuum with larger $\Gamma$ and the two black body components.
\cite{Becker2005ApJ} used a bulk Comptonization process to describe the energy transition from the kinetic energy of falling matters to the radiation. In such ``cold'' plasma, the kinetic energy of electrons is far more than their thermal energy, which results in a power-law shape without an energy cut, and the photon index will be larger than 2. This may end up in phase $0.8-1.0$ with a softer spectrum and higher $E_{\rm cut}$ because we cannot separate the cutoffpl component and the additional powerlaw component in the spectral fitting.

\begin{figure}
    \centering\includegraphics[width=0.49\textwidth]{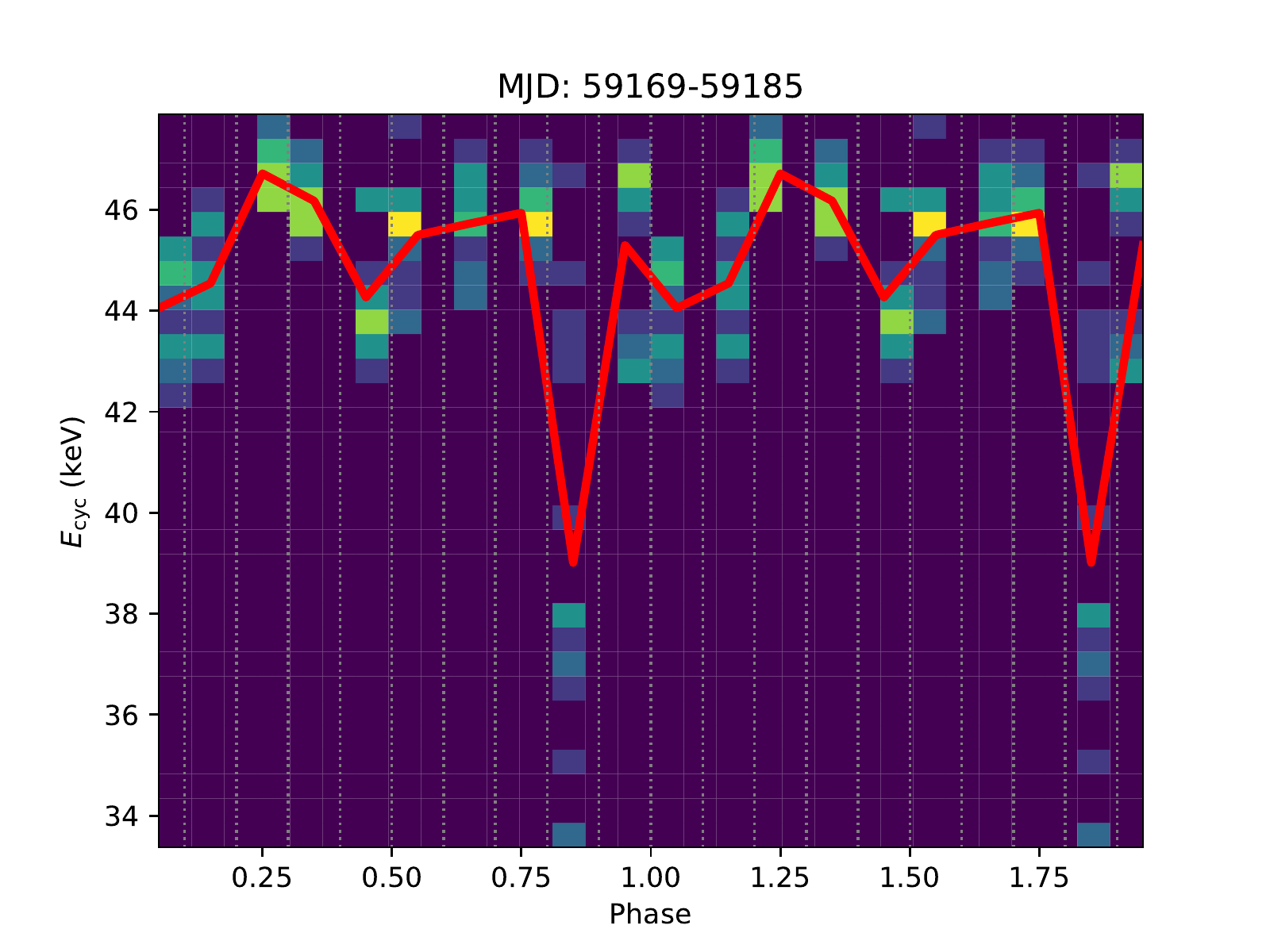}
    \centering\includegraphics[width=0.49\textwidth]{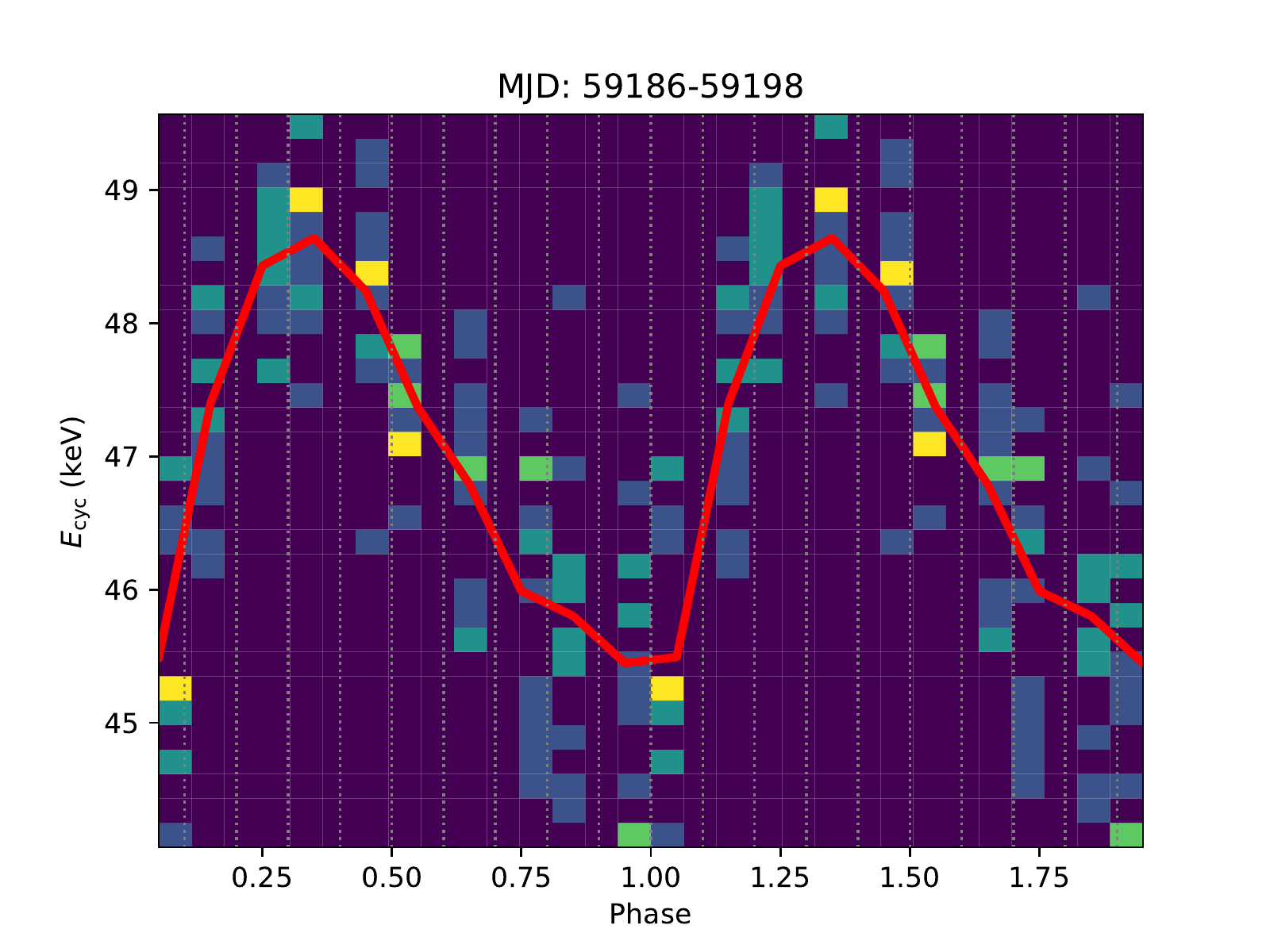}
    \centering\includegraphics[width=0.49\textwidth]{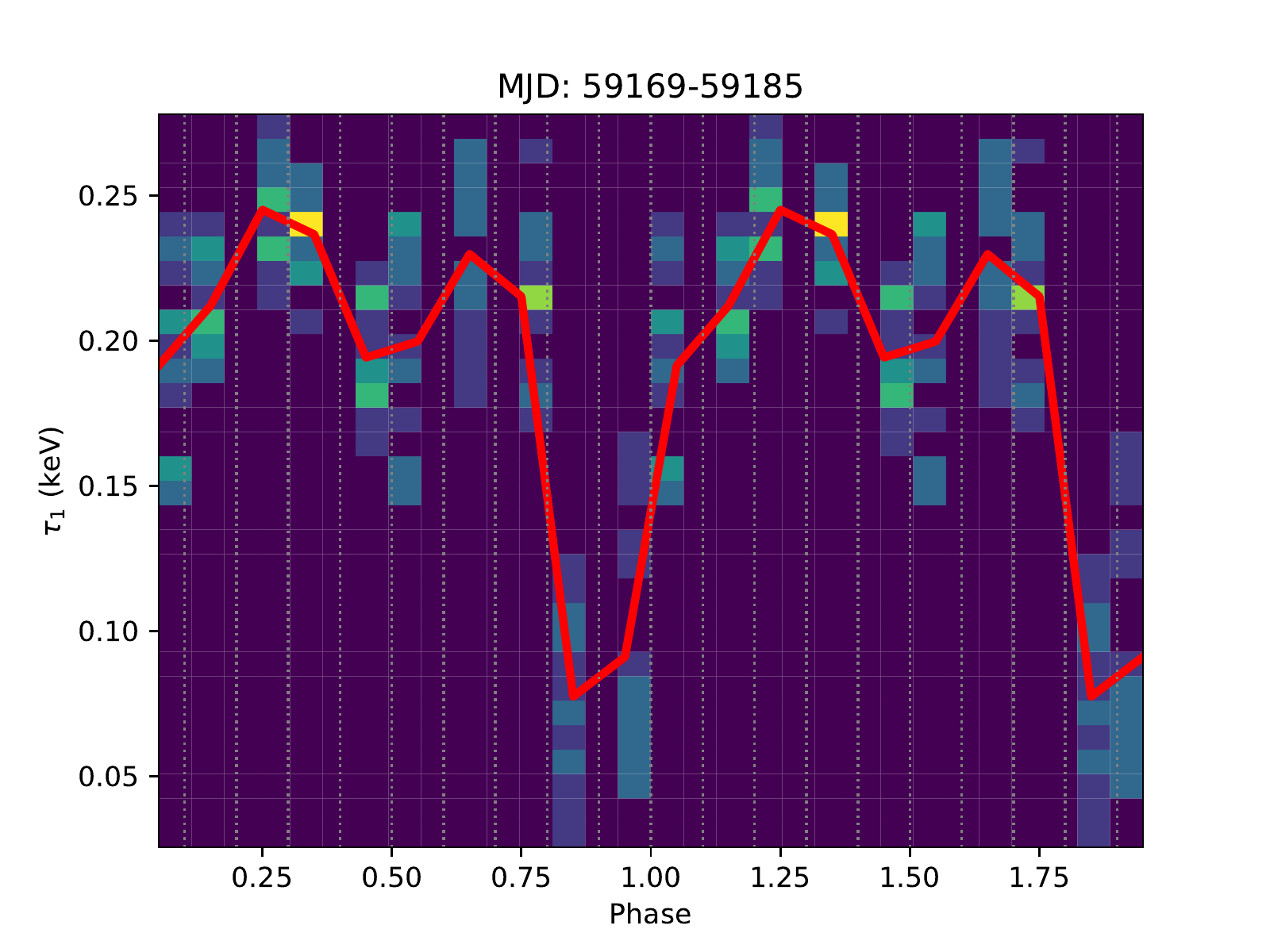}
    \centering\includegraphics[width=0.49\textwidth]{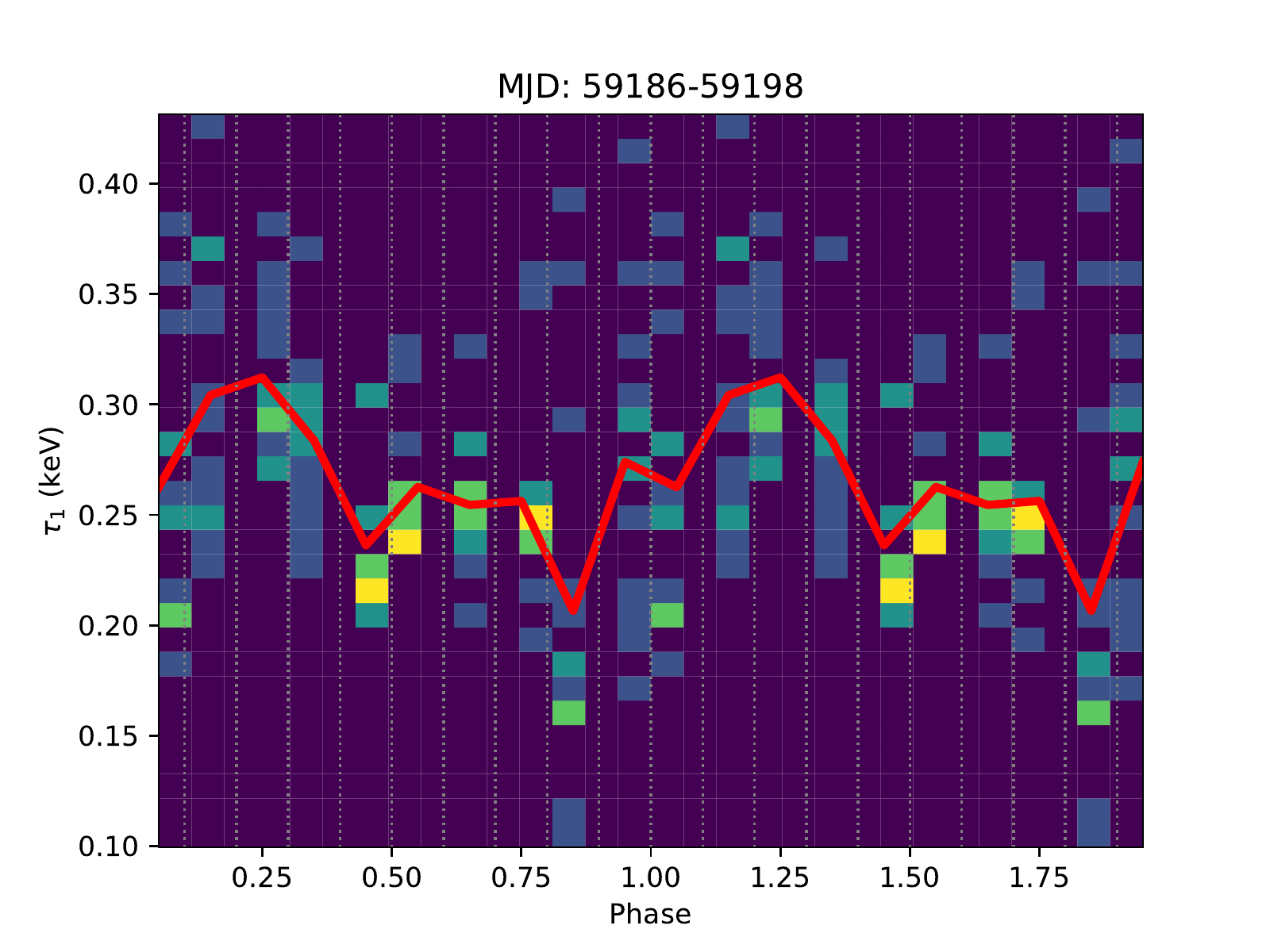}
    \caption{The left and right panels show the 2-D histogram of $E_{\rm cyc}$ (top) and $\tau_1$ (bottom), which represent the luminosity above (MJD $59169-59185$) and below the critical luminosity (MJD $59186-59198$). The red lines describe the average value of each phase. From the red lines of $E_{\rm cyc}$, we found a significant transition from a single peak (subcritical area) into a double peak (supercritical area).} 
    \label{45_distri}
\end{figure}

For the CRSF line at the fundamental, the 2-D distribution of line energy in Figure~\ref{2-D MPDs} splits into a double peak in phase $0.0-0.8$, and the dip in phase $0.8-1.0$ shows the shallowest absorption depth.
The left and right panels in Figure~\ref{45_distri} show the 2-D histogram of fundamental CRSF parameters which represent the luminosity above (MJD $59169-19185$) and below the critical luminosity (MJD $59186-19198$).
From the red lines in the two panels, we found a significant transition from a single peak (subcritical area) to a double peak (supercritical area).
In phase $0.8-1.0$, the first harmonic line dominates the absorption feature, while the fundamental line keeps low absorption depth.
In Figure~\ref{cor}, we compare the correlation between $\tau_1$ and $\tau_2$ from all observations. 
Pearson correlation coefficient $\rho=-0.03$ for phase $0.0-0.8$ and $\rho=-0.8$ for phase $0.8-1.0$ denoting a strong anti-correlation. 

For 1A~0535+262, we can identify a relatively narrow range of phases and luminosities where parameters of the CRSF line and non-thermal continuum appear to exhibit rapid changes.
Considering the value of the corresponding luminosity at $\sim 6\times10^{37}$ erg s$^{-1}$ is consistent with the $L_{\rm crit}$ in \cite{Kong2021}, and \cite{Mandal2022MNRAS} where the CRSF energy starts to show anti-correlation with luminosity we surmise the transition from sub- to supercritical accretion regime also responds to the transition in Figure~\ref{2-D MPDs}. 
In \cite{Becker2012}, the critical luminosity $L_{\rm crit}$ separates accretion column into two different patterns. 
When $L>L_{\rm crit}$, the deceleration of matters in the sinking area below the radiation-dominated shock is dominated by radiation pressure, and hence the height of the column increases with luminosity.
This results in the onset of a column with considerable height and optical depth along the magnetic field, and the radiation can only escape from the wall of the column, which is perpendicular to the magnetic field, forming the ``fan-beam'' pattern which associates double peaks of pulse profiles (\citealp{Davidson1973NPhS}). 
We note that the height of the column can reach a few km. Hence high energy photons that arise near the shock surface through bulk Comptonization (\citealp{Becker2005ApJ}) can be received effectively by the observer rather than be obscured or reflected by neutron stars (\citealp{Poutanen2013}), which makes it easier to see harmonics lines at higher energy during the supercritical regime. When $L<L_{\rm crit}$, the height of the column decrease with luminosity, and flow might be stopped primarily by Coulomb collisions (\citealp{Basko1976, Becker2012}). In this subcritical regime, the top of the column is finally located near the neutron star surface, where the plasma density is very high. At the same time, the photons can mainly escape from the top along the magnetic field as a ``pencil-beam'' pattern (\citealp{Burnard1991ApJ, Nelson1993ApJ}).   

Furthermore, through phase-resolved analysis, we confirm the presence of the significant 100 keV line in 1A~0535+262 reported by \cite{Kendziorra1994}, especially the absorption depth only assembles in a narrow phase.
We note that similar behavior was reported by \cite{Klochkov2008} close to the maximum of the 2006 giant outburst in EXO~2030+375, and they also found a lower fundamental line depth when a prominent first harmonic line is present.
We notice that the resonant scattering cross section at different levels shows different behaviors (\citealp{Harding1991}): for fundamental, $\sigma^1_{\rm res}\propto(1+\cos^2\theta)$, while for the first harmonic line, $\sigma^2_{\rm res}\propto(1+\cos^2\theta)\times \sin^2\theta$. 
The $\theta$ is the angle between the photon momentum and the
magnetic field. At different beaming pattern, the absorption depth for $n=1$ and $n=2$ is naturally different. So, the ``pencil-beam'' lead to  $\sigma^2_{\rm res}\ll\sigma^1_{\rm res}$.
Because of this, the harmonic line is naturally weak during subcritical regime, which is consistent with our results in Figure~\ref{2-D MPDs}.
During supercritical regime with ``fan-beam'', $\sigma^2_{\rm res}$ and $\sigma^1_{\rm res}$ can comparable. Therefore, it is reasonable to observe a stronger harmonic line in this luminosity range.
And we also notice that the harmonic absorption feature at higher energy sometimes makes the fundamental line more difficult to be detected. 
During a strong absorption of harmonic, the ``photon spawn'' effect and the ``superposition'' model have been suggested to account for a shallower fundamental line (\citealp{Nishimura2011, Nishimura2015}). 
For the ``photon spawn'' effect, the fundamental absorption feature can be filled up when the electrons, which have been excited by a high-energy incident photon to a higher Landau level, decay and generate photons with the energy around the fundamental line (\citealp{Schoenherr2007}). 
\cite{Nishimura2011, Nishimura2015} focuses on the influence of a superposition of a large number of lines arising from different heights and argues that it is expected to dominate over photon spawning. 
Hence, a shallower fundamental line and a deeper harmonic line can be observed in the spectrum. 
However, these models tend to expect a gradual change of the CRSF energy with phases and thus might not be applicable for our observations of the harmonic lines (see the figures in \cite{Schoenherr2007, Nishimura2011, Nishimura2015}).

We note that the pulse profiles have many complex components, which lead to another speculation.
We speculate the appearance of the first harmonic cyclotron line at a narrow phase during the supercritical area might be associated with a peculiar line-of-sight concerning the reflection and eclipse of the emission by the neutron star surface.
On one hand, the energy-dependent pulse profiles (left panel of Figure~\ref{fittings}) with multi-peaks structure or multi-dips below 20 keV reveal complicated emission patterns originating from the eclipse by NS surface (\citealp{Klochkov2008, Mushtukov2018MNRAS.474.5425M}).  
On the other hand, the reflection of the neutron surface for emission from the accretion column (\citealp{Poutanen2013, Caballero2011A&A...526A.131C}) can contribute a considerable fraction of the emission that arises from a reflecting halo because a significant part of the radiation from the column wall should be intercepted by the NS surface because of the relativistic beaming (\citealp{Kaminker1976,Lyubarskii1988}).
The reflection model implies minor phase variation of the CRSF energy because of the smooth change of the B-field strength over the surface.
This is consistent with our result of the fundamental line during the supercritical regime regardless of the phase $0.8-1.0$.
However, when energy goes above 20 keV, the shape of the pulse profiles is generally more straightforward in double peaks, which implies that the high-energy photons from the higher part of the column can not be affected by the neutron star surface in the direction of the line of sight.
We also notice that the phase where the spectrum shows a strong absorption feature at 100 keV does not locate at the peak of the pulse at high energy, which means that decreasing the numbers of high-energy photos by ellipse or reflection cannot explain the weak harmonic lines at other phases.
 
In Figure~\ref{fittings}, the pulse profile in $20-40$ keV gets the lowest in phase $0.8-1.0$, which means that the cap of the accretion column is facing the observer under a fan-beam pattern with strong relativistic beaming toward the NS surface. 
Because of this, most probably, the photons above 70 keV are from the opposite accretion column at phase $0.8-1.0$. 
And we monitor an additional component showing up in the pulse profile between the two peaks above 70 keV (see Figure~\ref{fittings} left panel).
Therefore, there is another possibility that the two lines come from different accretion columns.
In the context of having a dipole magnetic field, the two accretion columns on the magnetic pole are symmetric and on the two sides of the neutron star surface.
When one pole faces the observer, the other locates the antipodal position. 
Under this assumption, the high-energy photons born out of a specific height of the accretion column can create another emission pattern due to the gravitational bending effect while the accretion column is on the other side of the neutron star as a ``anti-pencil'': the photons can be focused and become visible within a narrow phase (\citealp{Sasaki2010, Mushtukov2018MNRAS.474.5425M, Molkov2019}). 
This new anti-pencil pattern can contribute to high-energy photons subjected to CRSF absorption at 100 keV.

\section{Conclusions}
In this paper, the phase-resolved spectral analysis is used to explore the variation of spectral parameters with luminosity and phases.
We discover the elaborate distribution and evolution of the parameters from CRSF and other spectral components below and above the critical luminosity. 
Above critical luminosity, our results show the fundamental line dominates almost all periods but it gets weaker and lower with the appearance of harmonics which can only concentrate in a narrow phase interval. 
Here, we give a possible surmise that the strict pulse phase dependence of the absorption depth can be attributed to the ``anti-pencil beam'' pattern. 
Future observation and theoretical modeling will investigate the probability of having such a pattern.

\acknowledgments
This work used data from the Insight-HXMT mission, a project funded by China National Space Administration (CNSA) and the Chinese Academy of Sciences (CAS).
This work is supported by the National Key R\&D Program of China (2021YFA0718500) and the National Natural Science Foundation of China under grants U1838201, U2038101, 11473027, U1838202, 11733009, 12173103, U1838104, U1938101, U1938103 and Guangdong Major Project of Basic and Applied Basic Research (Grant No. 2019B030302001). MO acknowledges support by the Italian Space Agency under grant ASI-INAF n.~2017--14--H.0.

\bibliography{A0535.bbl}
\bibliographystyle{aasjournal}

\end{document}